\documentclass[number,sort&compress,5p]{elsarticle}
\pdfoutput=1
\usepackage{amsmath,amssymb}
\usepackage{mathptmx}
\usepackage{graphicx}
\usepackage{hyperref}
\newcommand\blfootnote[1]{%
  \begingroup
  \renewcommand\thefootnote{}\footnote{#1}%
  \addtocounter{footnote}{-1}%
  \endgroup
}
\newcommand{\appendices}{
  \renewcommand{\thesection}{\Alph{section}}
  \setcounter{section}{0}
  \renewcommand{\theequation}{\Alph{section}-\arabic{equation}}
  \setcounter{equation}{0}
  \renewcommand{\thefigure}{\Alph{section}-\arabic{figure}}
  \setcounter{figure}{0}
  \renewcommand{\thetable}{\Alph{section}-\arabic{table}}
  \setcounter{table}{0}
}
\def\mathbi#1{\textbf{\em #1}}
\renewcommand{\vec}[1]{\boldsymbol{\mathrm{#1}}}

\begin{document}
\title{Modified Gravity (MOG) fits to observed radial acceleration of
  SPARC galaxies}

\author[PI]{M. A. Green}
\ead{mgreen@perimeterinstitute.ca}
\author[PI,UW]{J. W. Moffat}
\ead{jmoffat@perimeterinstitute.ca}
\address[PI]{Perimeter Institute for Theoretical Physics,
  Waterloo ON N2L 2Y5, Canada}

\address[UW]{Department of Physics and Astronomy, University of
  Waterloo, Waterloo ON N2L 3G1, Canada}

\date{}

\begin{abstract}\noindent The equation of motion in the generally covariant
  modified gravity (MOG) theory leads, for weak gravitational fields
  and non-relativistic motion, to a modification of Newton's
  gravitational acceleration law.  In addition to the metric
  $g_{\mu\nu}$, MOG has a vector field $\phi_\mu$ that couples with
  gravitational strength to all baryonic matter.  The gravitational
  coupling strength is determined by the MOG parameter $\alpha$, while
  parameter $\mu$ is the small effective mass of $\phi_\mu$. The MOG
  acceleration law has been demonstrated to fit a wide range of
  galaxies, galaxy clusters and the Bullet Cluster and Train Wreck
  Cluster mergers.  For the SPARC sample of rotationally supported
  spiral and irregular galaxies, McGaugh et al.\ \cite{McGaugh} (MLS)
  have found a radial acceleration relation (RAR) that relates
  accelerations derived from galaxy rotation curves to Newtonian
  accelerations derived from galaxy mass models.  Using the same SPARC
  galaxy data, mass models independently derived from that data, and
  MOG parameters $\alpha$ and $\mu$ that run with galaxy mass, we
  demonstrate that adjusting galaxy parameters within $\pm 1$-sigma
  bounds can yield MOG predictions consistent with the given
  rotational velocity data.  Moreover, the same adjusted parameters
  yield a good fit to the RAR of MLS, with the RAR parameter
  $a_0=(5.4\pm .3)\times 10^{-11}\,{\rm m/s^2}$.
\end{abstract}

\maketitle

\section{Introduction}

\blfootnote{\noindent\copyright 2019. This manuscript version is made available
under the CC-BY-NC-ND 4.0 license\\\url{http://creativecommons.org/licenses/by-nc-nd/4.0/}}
The pioneering research by Zwicky~\cite{Zwicky} and Vera Rubin and
collaborators~\cite{Rubin1,Rubin2} showed an apparent discrepancy
between the dynamics of galaxies and galaxy clusters and the
predictions of Newtonian and Einstein gravity.  The dynamical masses
inferred from observations of galaxies and clusters were found to
exceed the baryon mass in these systems.  The existence of mysterious
dark matter, forming halos around galaxies, was proposed to resolve
the discrepancy.  To date, there is no convincing evidence in deep
underground laboratory experiments, such as LUX~\cite{LUX} and
Panda-X~\cite{PandaX}, astrophysical observations, or LHC experiments
to support the existence of exotic dark matter particles.

An alternative way to explain the observed dynamics of galaxies and
clusters is to adopt a modified theory of gravitation.  Modified
Newtonian dynamics (MOND)~\cite{Milgrom1,Milgrom2} uses an empirical
non-relativistic formula for gravitational acceleration to fit
observed galaxy velocity profiles.  MOND combines a characteristic
minimum acceleration scale $a_0=1.2\times 10^{-10}\,{\rm m/s^2}$ with
empirical choice from many possible interpolating functions.  But it
is not based on a covariant fundamental action principle and MOND
fails to fit clusters and cosmology without dark matter.

A more satisfactory modified gravitational theory would fulfil the
following requirements:

\begin{itemize}
  \setlength{\itemsep}{3pt}
\item general covariance and local Lorentz invariance
\item classical causal locality
\item equivalence principle
\item contains General Relativity (GR) in a well-defined limit.
\end{itemize}

\noindent Modified gravity (MOG), originally called
Scalar-Tensor-Vector-Gravity (STVG), satisfies these four
requirements~\cite{Moffat1}.  MOG can explain the dynamics of
galaxies, clusters and the large-scale structure of the universe
without needing to add dark matter.  Using phase-space analysis, it was
shown in \cite{Jamali} that MOG has a viable sequence of cosmological
epochs.  MOG has been shown to fit a large number of galaxy rotation
curves~\cite{MoffatRahvar1}, globular cluster velocity dispersions
\cite{MoffatToth2} and lensing observations \cite{MoffatRahvarToth},
and ultra-diffuse galaxy NGC 1052-DF2 \cite{MoffatToth18}, satisfy the
Tully-Fisher relation~\cite{TullyFisher,Verheijen}, successfully
describe the dynamics of clusters~\cite{MoffatRahvar2} and merging
clusters, such as the Bullet Cluster and the Train Wreck Cluster Abell
520~\cite{BrownsteinMoffat,IsraelMoffat}, and model the growth of
structure to fit the observed matter power spectrum at present
\cite{MoffatToth3,Moffat06,Moffat15,Jamali18}.

The LIGO/Virgo observatory detection of the merging neutron stars
gravitational wave event $GW170817$, together with the optical
detection $GRB170817A$, has determined that gravitational waves move
with the speed of light to one part in $10^{15}$.  MOG is also
compatible with this observational result~\cite{MoffatGreenToth}.

We report here on an investigation of the match between gravitational
accelerations calculated using MOG and accelerations inferred from
rotational velocities for a diverse set of galaxies.  McGaugh et
al.~\cite{McGaugh} (MLS) have demonstrated an empirical radial
acceleration relation (RAR) between predicted Newtonian accelerations
and accelerations inferred from observations of 153 galaxies in the
Spitzer Photometry and Accurate Rotation Curves (SPARC) dataset.  A
characteristic deviation from Newtonian gravitational acceleration is
evident for a wide range of galaxy masses and types.  In the
following, we will demonstrate the ability of MOG to fit accelerations
inferred from the same SPARC galaxy data, with suitable allowances for
the uncertainties of galaxy parameters.

For each SPARC galaxy, data made available by McGaugh et
al.\footnote{Data for the SPARC galaxies was obtained from
  \url{http://astroweb.cwru.edu/SPARC/}.} includes the galaxy
distance, inclination, $3.6~\mu\mathrm{m}$ surface photometry, total
luminosity, HI mass and radius, and rotational velocity profiles.  The
provided data also lists the inferred contributions of bulge, stellar
disk and gas components to Newtonian rotational velocity at radii
corresponding to the rotational velocity measurements.  Calculation of
these velocity contributions clearly required models for the
3-dimensional mass distributions of bulge, disk and gas within the
galaxy; but details were not provided for the mass models used in MLS.
Mass models are also needed to calculate MOG accelerations.  We have
used the surface photometry and other data listed above to develop
the needed galaxy mass models.

The next Section introduces the MOG acceleration law for
non-relativistic systems with weak fields.  It is followed by a
summary of analyses of SPARC galaxy data and fits to the RAR reported
in MLS and subsequent studies.  In Section \ref{sec:SPARC-MOG} we
describe our independent analysis of the SPARC data, our own fit to
the RAR, comparison of MOG predictions to observed accelerations, and
adjustment of uncertain galaxy parameters to improve MOG fits.  We
discuss the significance of the many assumptions and uncertainties
involved in analysis and fitting of the galaxy data to empirical or
theoretical models and draw conclusions.

\section{MOG Acceleration Law}
Like GR, MOG is a classical theory of gravitation based on a fully
covariant action principle and corresponding field equations.  In
addition to the spacetime metric $g_{\mu\nu}$, the MOG formalism
introduced in~\cite{Moffat1}, and summarized in
Appendix~\ref{app.MOGsummary}, has a massive vector field $\phi_\mu$
and non-negative scalar fields $\alpha$ and $\mu$. The MOG
gravitational coupling strength is $G=G_N(1+\alpha)$, where $G_N$ is
Newton's gravitational constant.  MOG reduces to GR in the limit
$\alpha\to 0$.

The equation of motion for a massive test particle in MOG has the
covariant form~\cite{Moffat1}:
\begin{equation}
\label{eqMotion}
m\biggl(\frac{du^\mu}{ds}+{\Gamma^\mu}_{\alpha\beta}u^\alpha
u^\beta\biggr)= q_g{B^\mu}_\nu u^\nu,
\end{equation}
where $u^\mu=dx^\mu/ds$ with $s$ the proper time along the particle
trajectory, ${\Gamma^\mu}_{\alpha\beta}$ denote the Christoffel
symbols, and $B_{\mu\nu}=\partial_\mu\phi_\nu-\partial_\nu\phi_\mu$.
Moreover, $m$ and $q_g$ denote the test particle mass $m$ and
gravitational charge $q_g=\sqrt{\alpha G_N}m$, respectively.  We note
that for $q_g/m=\sqrt{\alpha G_N}$ the equation of motion for a
massive test particle (\ref{eqMotion}) {\it satisfies the (weak)
  equivalence principle}, leading to the free fall of particles in a
homogeneous gravitational field, although the free-falling particles
do not follow geodesics.

In the weak field region, $r\gg 2GM$, surrounding a stationary mass
$M$ centred at $r=0$ the spherically symmetric field $\phi_\mu$, with
effective mass $\mu$, is well approximated by the Yukawa potential:
\begin{equation}
  \phi_0=-Q_g\frac{\exp(-\mu r)}{r},~~~~\phi_i=0,~i=1..3\,,
\end{equation}
where $Q_g=\sqrt{\alpha G_N}M$ is the gravitational charge of $M$.
The radial equation of motion of a non-relativistic test particle,
with mass $m$ and at radius $r$, in the field of $M$ is then given by
\begin{equation}
  \label{Eq:MOGweakEOM}
  \frac{d^2r}{dt^2}+\frac{GM}{r^2}=
  \frac{q_gQ_g}{m}\frac{\exp(-\mu r)}{r^2}(1+\mu r).
\end{equation}
The mass $\mu$ is tiny\,---\,comparable to the experimental bound on
the mass of the photon\,---\,giving a range $\mu^{-1}$ of the
repulsive exponential term the same order of magnitude as the size of
a galaxy.  Since $ q_gQ_g/m=\alpha G_NM$, the modified Newtonian
acceleration law for a point particle can be written
as~\cite{Moffat1}:
\begin{equation}
\label{MOGaccelerationlaw}
a_{\rm MOG}(r)=-\frac{G_NM}{r^2}[1+\alpha-\alpha\exp(-\mu r)(1+\mu r)].
\end{equation}
This reduces to Newton's gravitational acceleration in the limit
$\mu r\ll 1$.

In the limit that $r\rightarrow\infty$, we get from
(\ref{MOGaccelerationlaw}) for approximately constant $\alpha$ and
$\mu$:
\begin{equation}
\label{AsymptoticMOG}
a(r)\approx -\frac{G_N(1+\alpha)M}{r^2}.
\end{equation}
MOG has a Newtonian-Kepler behaviour for large $r$ with enhanced
gravitational strength $G=G_N(1+\alpha)$. The transition from
Newtonian acceleration behavior for small $r$ to non-New-tonian
behaviour for intermediate values of $r$ is due to the repulsive
Yukawa contribution in (\ref{MOGaccelerationlaw}). This can also
result in the circular orbital rotation velocity $v_c$ having a
maximum value in the transition region.  The prediction of
Newton-Kepler-like rotation curves at large $r$, with $G>G_N$,
distinguishes MOG from the MOND asymptotic behavior.  A first
verification of the MOG asymptotic behavior as $r\rightarrow\infty$
has been obtained from a fitting of Milky Way galaxy rotation curves
with a distance scale $R\sim 200$ kpc~\cite{MoffatToth1}.

For a distributed baryonic matter source, the MOG (weak field)
acceleration law becomes \cite{MoffatRahvar1}:
\begin{eqnarray}
  \label{accelerationlaw2}
  &\hspace*{-1.4in}{\mathbi a}_{\rm MOG}({\vec x})=-G_N\int d^3{\vec x}'
  \frac{\rho_{\rm bar}({\vec x}')({\vec x}-{\vec x}')}{|{\vec x}-{\vec x}'|^3}\nonumber\\
  &\hspace*{.3in}[1+\alpha-\alpha\exp(-\mu|{\vec x}-{\vec x}'|)(1+\mu|{\vec x}-{\vec x}'|)]\,,
\end{eqnarray}
where $\rho_{\rm bar}$ is the total baryon mass density.  In
the following, $a_{\rm MOG}$, $a_{\rm obs}$ and $a_{\rm bar}$ will
denote the magnitudes of the accelerations indicated by the suffix.

Equations (\ref{Eq:MOGweakEOM}) to (\ref{accelerationlaw2}) were
derived using the assumption that variations of $\alpha$ and $\mu$ are
ignorably small within the spacetime region being considered.  In this
paper we adopt a simplified version of MOG that formalizes this
assumption.  Instead of Eq.~(\ref{Eq:SS}) we set $S_S=0$, and treat
$\alpha$ and $\mu$ as parameters that run, taking effectively constant
values that depend on the scale of the system under investigation and
the spatial resolution with which it is observed.  This is analogous
to the running of masses and coupling parameters associated with
renormalization group (RG) flow when a condensed matter or particle
physics system is observed at different scales.

\section{
  The Radial Acceleration Relation}

Data for 153 galaxies from the SPARC sample of galaxies \cite{Lelli}
was used in MLS to demonstrate correlation between the Newtonian
acceleration $a_{\rm bar}$, due to the apparent baryonic matter, and
the acceleration $a_{\rm obs}$ inferred from galaxy rotational
velocity curves.  Fig.~3 of MLS shows, as a histogram of $2693$
individual data points for 153 different galaxies, results that
demonstrate obvious correlation between $g_{\rm obs}$ (our
$a_{\rm obs}$) and $g_{\rm bar}$ (our $a_{\rm bar}$).\footnote{
  Exclusion of rotational velocity data points with $\le 10$\%
  uncertainty means that only 147 galaxies actually contribute to
  Fig.\ 3 of MLS.  The present work excludes points with $<10$\%
  uncertainty, thus keeping 7 more data points and increasing the
  number of galaxies to 149.}  In MLS, a RAR that fits the empirical
data is given by
\begin{equation}
  \label{Empiricalacc}
  a_{\rm obs}={\cal F}(a_{\rm bar})=\frac{a_{\rm bar}}
  {1-\exp(-\sqrt{a_{\rm bar}/a_0})},
\end{equation}
where
$a_0=(1.20\pm 0.02\,({\rm random})\pm 0.24\,({\rm systematic}))\times
10^{-10}$ ${\rm m}\,{\rm sec}^{-2}$. The random error is a $1\sigma$
value, while the $20\,\%$ systematic uncertainty is attributed to
normalization uncertainty of the stellar mass-to-light ratio.
Eq.~(\ref{Empiricalacc}) contains the one critical acceleration scale
$a_0$; for $a_{\rm bar} \gg a_0$ it gives
$a_{\rm obs}\approx a_{\rm bar}$ and for $a_{\rm bar} \ll a_0$ it
gives $a_{\rm obs} \approx \sqrt{a_{\rm bar}\,a_0}$.

The results of MLS were obtained using best estimated values for
galaxy inclination, $i$, and galaxy distance, $D$.  Mass models were
derived from $3.6~\mu\mathrm{m}$ images and HI profiles.  The HI data
was compiled from other studies and made available in the SPARC
database, without uncertainties, as total HI mass $M_{\rm HI}$ and
radius $R_{\rm HI}$ where HI surface density (corrected to face-on)
reaches $1~M_\odot {\rm pc}^{-2}$.  Generic functional forms were
assumed for bulge and stellar disk components; but algorithmic details
of the decomposition into two or three (including bulge) components
and transformation from the inclined line-of-sight (LoS) to face-on
(FO) were not provided.  The total mass of gas was assumed to be
$1.33\,M_{\rm HI}$.\footnote{The contributions, $v_{\rm gas}$, of each
  galaxy's gas to rotational velocity reflect the compiled HI
  profiles, which were not included with the SPARC data and not used
  in the present work.}  Disk and bulge mass-to-light ratios were
assumed to have the fixed values
$\Upsilon_{\rm disk}=\Upsilon_\star=0.5~M_\odot/L_\odot$,
$\Upsilon_{\rm bul}=1.4\,\Upsilon_{\rm disk}$.  This is supported by
analysis presented in \cite{Lelli et al.(2017)}, which showed the RARs
for individual galaxies and investigated the effect of choosing
different fixed values for $\Upsilon_{\rm bul}$ and
$\Upsilon_{\rm disk}$.  Plots in \cite{Lelli et al.(2017)} show that
for many individual galaxies, especially those with
$M_{\rm bar}<3\times 10^{10}~{\rm M_\odot}$, the observed
accelerations deviate greatly from the functional form
(\ref{Empiricalacc}) with a fixed value of $a_0$.

Choosing fixed $a_0=1.20\times 10^{-10}\, {\rm m}\,{\rm sec}^{-2}$, Li
et al.\ \cite{Li et al.(2018)} fit the RAR (\ref{Empiricalacc}) to
individual galaxies by marginalizing over $\Upsilon_\star$, $i$ and
$D$, with Gaussian priors whose widths match estimated observational
errors.  They did similar fits with $a_0$ added to the list of
adjustable parameters; separate analyses were done assuming a prior
for $a_0$ that is flat from 0 to $10^{-9}\, {\rm m}\,{\rm sec}^{-2}$,
and a Gaussian prior with
$a_0=(1.20\pm 0.02)\times 10^{-10}\, {\rm m}\,{\rm sec}^{-2}$.  As one
should expect, use of the best fit parameters yields histograms with
much less dispersion about the RAR curve than in Fig.~3 of MLS.

Fig.\ 7 of \cite{Li et al.(2018)} shows that, in the above case of
$a_0$ with a flat prior, the distribution of optimal $a_0$ values has
a full width of about an order of magnitude.  Rodrigues et al.\
\cite{Rodrigues2018}, marginalizing over
$\Upsilon_{\rm disk}$, $\Upsilon_{\rm bul}$, $D$ and $a_0$ but not $i$
found a similarly broad distribution.  From the significant variation
and uncertainty of fitted $a_0$ values, they conclude that the
``emergent acceleration'' obtained by combining the results from
individual galaxies ``cannot be considered a fundamental
acceleration''.

\section{Analysis of SPARC data with MOG}\label{sec:SPARC-MOG}

The ``mass models'' made available by \mbox{McGaugh} et al.\ for the
SPARC galaxies specify the face-on surface density only at radii for
which velocity measurements are available.  Since calculation of MOG
accelerations, using Eq.~(\ref{accelerationlaw2}), requires knowledge
of $\rho_{\rm bar}$ at all $(r,z)$ within a galaxy, we used the given
data, excluding rotational velocities, to independently develop the
needed mass models.

Decomposition of the $3.6~\mu\mathrm{m}$ photometry observations into
bulge and disk surface brightness radial profiles is described in
\cite{Lelli}.  Bulges were identified for 32 galaxies; the remainder
were assumed to have disks only.  Our independent analysis of the
photometry data, to determine and characterize bulge and disk
components, identified bulges for 57 of 149 galaxies.  Exponentials
were fitted to the outer disk and used to extrapolate to larger radii.
Bulges were assumed to be spherical.  Stellar disks were assumed to
have a ${\rm sech}^2$ vertical mass distribution \cite{Spitzer,
  Allaert} with scale height $z_{\rm d}=0.196\,{R_{\rm d}}^{0.633}$,
where $R_{\rm d}$ is the given scale length in kpc \cite{Bershady}.

We converted the observed stellar disk surface brightness profiles
$\Sigma_{\rm disk,obs}(r)$ to equivalent face-on profiles
$\Sigma_{\rm disk,FO}(r)$ by numerically integrating along the
line-of-sight, with inclination $i$, through a provisional stellar
density model $\rho_{\rm disk}(r,z)$ to determine the corresponding
model line-of-sight surface brightness $\Sigma_{\rm disk,LoS}(r)$.
Keeping the $z$ dependence fixed, $\rho_{\rm disk}$ was then
iteratively adjusted to make the ratio
$\Sigma_{\rm disk,LoS}(r)/\Sigma_{\rm disk,obs}(r)$ converge to
1.\footnote{The iteration was started by treating the observed profile
  $\Sigma_{\rm disk,obs}(r)$ as face-on, with vertical mass
  distribution as described above.}  This gave the 3-d stellar disk
density.  A similar iterative process was used to determine the bulge
radial density $\rho_{\rm bul}$ from the bulge surface brightness
$\Sigma_{\rm bul,obs}(r)$.  Plots of the bulge-disk decompositions and
derived face-on models are provided in the Supplementary Material.

Instead of the fixed ratio 1.33, we chose gas disk models that reflect
a variation of $\eta=M_{\rm gas}/M_{\rm HI}$ with galaxy morphological
type, where $M_{\rm gas}$ is the total gas mass.  This allows for
typical abundances of gases other than HI and He, for which there was
no data or modeling allowance in the analysis of MLS.  We also
incorporated depressions in the central regions
\cite{2014MNRAS.441.2159W}.  Our gas density models do not consider
the radial variation of the ratio of HI to other gases, nor do they
have the benefit of observed HI profiles.  Actual total gas density
profiles are likely to have large statistical variations (as much as 1
dex for early-type galaxies) about these models
\cite{2014AA...564A..66B}.  For the ratio $\eta$ we used a relation
from \cite{McGaugh97} based on the numerical morphological type $T$:%
\begin{equation}
  \eta(T) = \left\{\genfrac{}{}{0pt}{}{1.4\,(4.7-0.8\,T+0.043\,T^2)\,,~~T<9\,,}
  {1.4\,,~~T\ge 9\,.\hfill}\right.\label{eq.eta}
\end{equation}
These values range from $\eta(0)=6.58$ to $\eta(9)=1.4$, and increase
the average total mass of the 153 studied galaxies by 15 percent when
compared to masses calculated with $\eta=1.33$.  The gas disk was
assumed to have the same effective scale height as the stellar disk.
A radial exponential form with scale length $r_g$ was assumed at large
radius, but this was depressed in the central region by the factor%
\begin{equation}
  f(r)=\left[1-\frac{1}{\eta(T)}\left(1+\frac{r_g}{R_{\rm d}}\right)^{-2}\right]
  \left(1-\frac{1}{\eta(T)}\exp(-r/R_{\rm d})\right)\,.
\end{equation}
The factor in square brackets is for normalization: the central gas
density and $r_g$ were chosen to yield the total gas mass
$M_{\rm gas}=\eta(T)M_{\rm HI}$ and the given $R_{\rm HI}$ at modeled
HI surface density of $1~{\rm M_\odot/pc^2}$.

\begin{figure}[t!]
  \centering \includegraphics[width=3.25in]{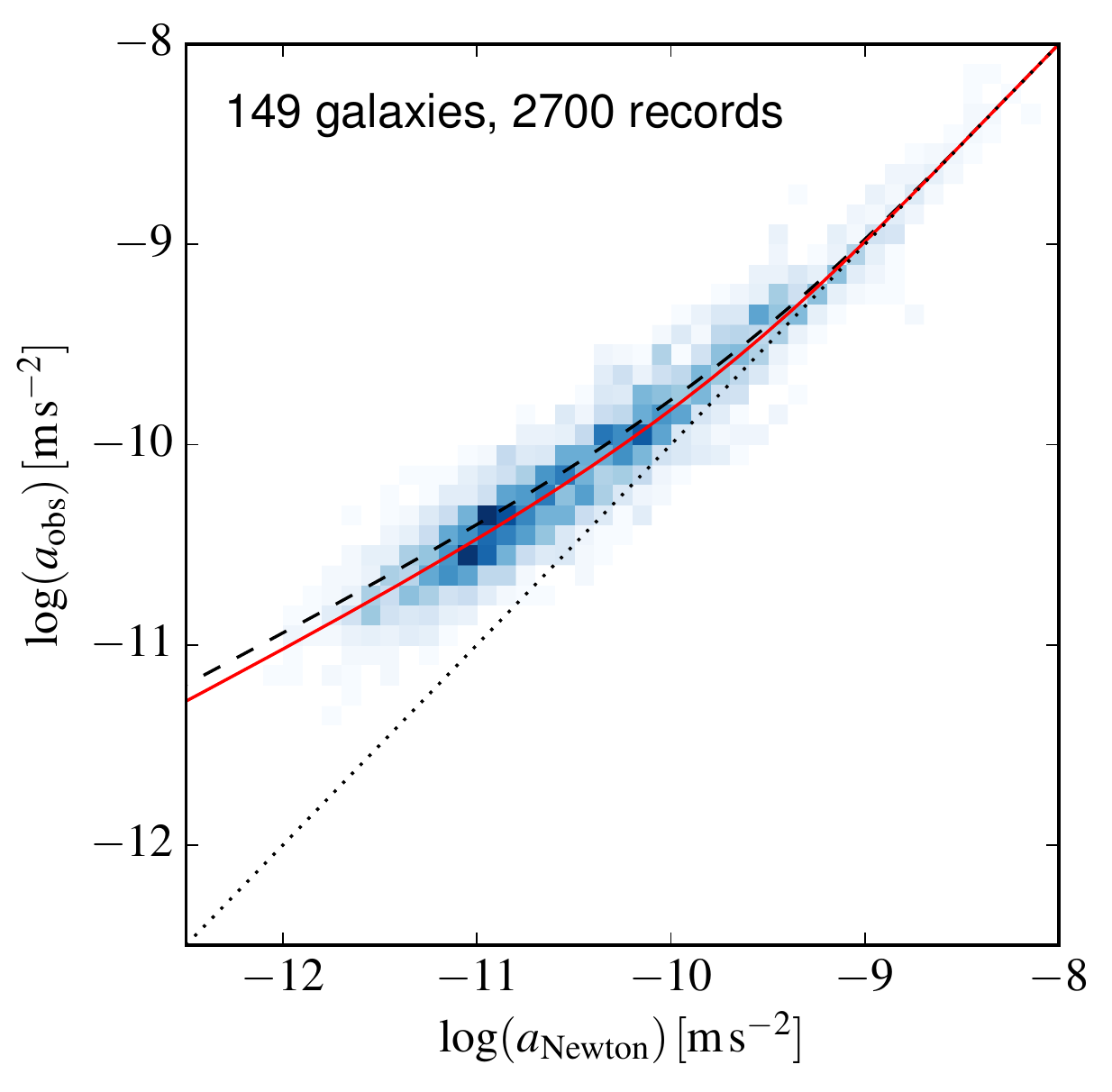}
  \caption{Relation of centripetal acceleration $a_{\rm obs}$ to
    Newtonian acceleration $a_{\rm Newton}$, based on our new mass
    models.  The dashed black curve is the RAR (\ref{Empiricalacc}),
    with $a_{\rm bar}=a_{\rm Newton}$ and
    $a_0=1.2\times 10^{-10}\,{\rm m/s^2}$.  The solid red curve is the
    RAR using $a_0=(8.1\pm .6)\times 10^{-11}\,{\rm m/s^2}$, which is
    the mean $\pm 1\,\sigma$ of the best fit values for individual
    galaxies.  The fit for each galaxy weights data points inversely
    with both the fractional uncertainties of $a_\mathrm{obs}$ and the
    local density of points in the radial direction.  Closely spaced
    $v_\mathrm{obs}$ records are thus prevented from dominating over
    more widely spaced records.  Similarly weighted residuals relative
    to the RAR curve have a width $\sigma=0.14$~dex, compared with
    $\sigma=0.11$~dex in Fig.\ 3 of MLS.}\label{fig.histNewton}
\end{figure}

The relation of Newtonian gravitational acceleration $a_{\rm Newton}$
(calculated using our new mass models) to the centripetal acceleration
$a_{\rm obs}$ (inferred from observed rotational velocities,
$v_{\rm obs}$) is shown in Fig.~\ref{fig.histNewton}.  The histogram
is similar to that of Fig.~3 of MLS, with about 30 percent less
discrepancy between $a_{\rm Newton}$ and $a_{\rm obs}$\,.  In
Appendix~\ref{app.appAlt} we show that the reduced discrepancy arises
from the increased mass of gas in our models.  The Supplementary
Material includes plots for each galaxy of observed and calculated
Newtonian rotation curves (with stellar and gas components).  Whether
the mass models of McGaugh et al.\ or our mass models are more
representative of reality will require more information to decide.
Qualitatively they give similar results, so use of our mass models to
calculate MOG accelerations seems reasonable.

\begin{figure}[t!]
  \centering \includegraphics[width=3.25in]{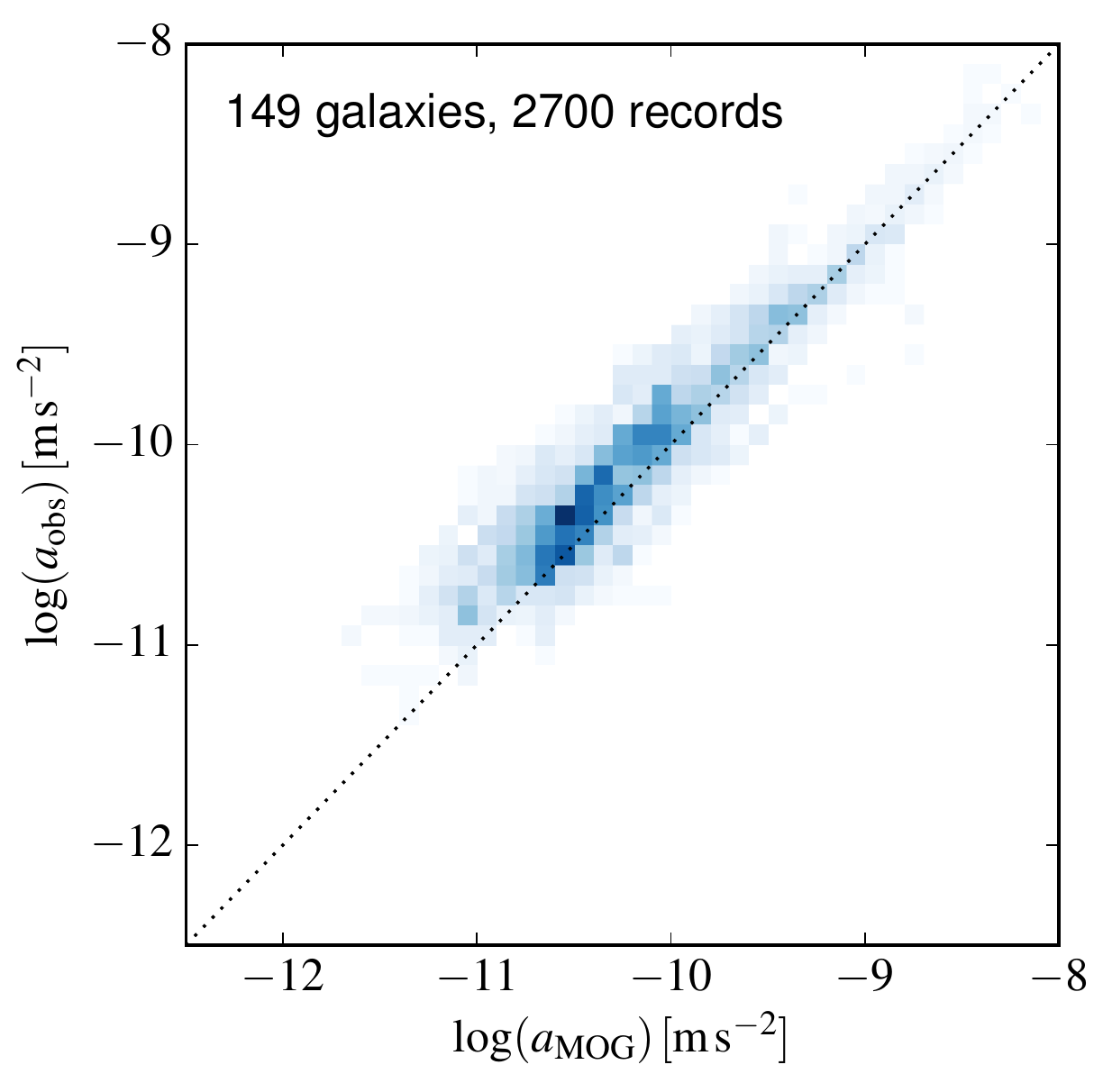}
  \caption{Relation of centripetal acceleration $a_{\rm obs}$ to MOG
    acceleration $a_{\rm MOG}$, using the same galaxy parameters as
    Fig.\ \ref{fig.histNewton}.  The mean
    $\langle \log(a_{\rm obs}/a_{\rm MOG})\rangle =0.13$~dex, and the
    residuals have width $\sigma=0.21$~dex.}\label{fig.histMOG}
\end{figure}

For a given mass density model $\rho_{\rm bar}(r,z)$, MOG
accelerations can be computed using Eq.~(\ref{accelerationlaw2}) and
numerical integration.  For the running MOG parameters $\mu$ and
$\alpha$ we have used the functional forms~\cite{MoffatToth(2009)}:
\begin{equation}
  \mu(M)=\frac{D_0}{\sqrt{M}}\,,\hspace*{.2in}
  \alpha(M)=\alpha_\infty\frac{M}{(\sqrt{M}+E_0)^2}\,,\label{eq.alpha-mu}
\end{equation}
where $M$ is the total mass of the galaxy, $\alpha_\infty=10$,
$D_0=6.25\times 10^3~\sqrt{M_\odot}/{\rm kpc}$ and
$E_0=2.5\times 10^4~\sqrt{M_\odot}$.  Eqs.~(\ref{eq.alpha-mu}) were
derived from a point source solution of the MOG field equations and
should thus be considered only approximately valid for mass
distributed as in a galaxy.  The given values of $\alpha_\infty$,
$D_0$ and $E_0$, determined by fitting to a small sample of galaxies,
are subject to adjustment as warranted by additional data; but they
are used here without change.  Comparison of the calculated MOG
gravitational acceleration $a_{\rm MOG}$ with the centripetal
acceleration $a_{\rm obs}$ is shown in Fig.~\ref{fig.histMOG}.

\begin{figure}[t!]
  \centering \includegraphics[width=3.25in]{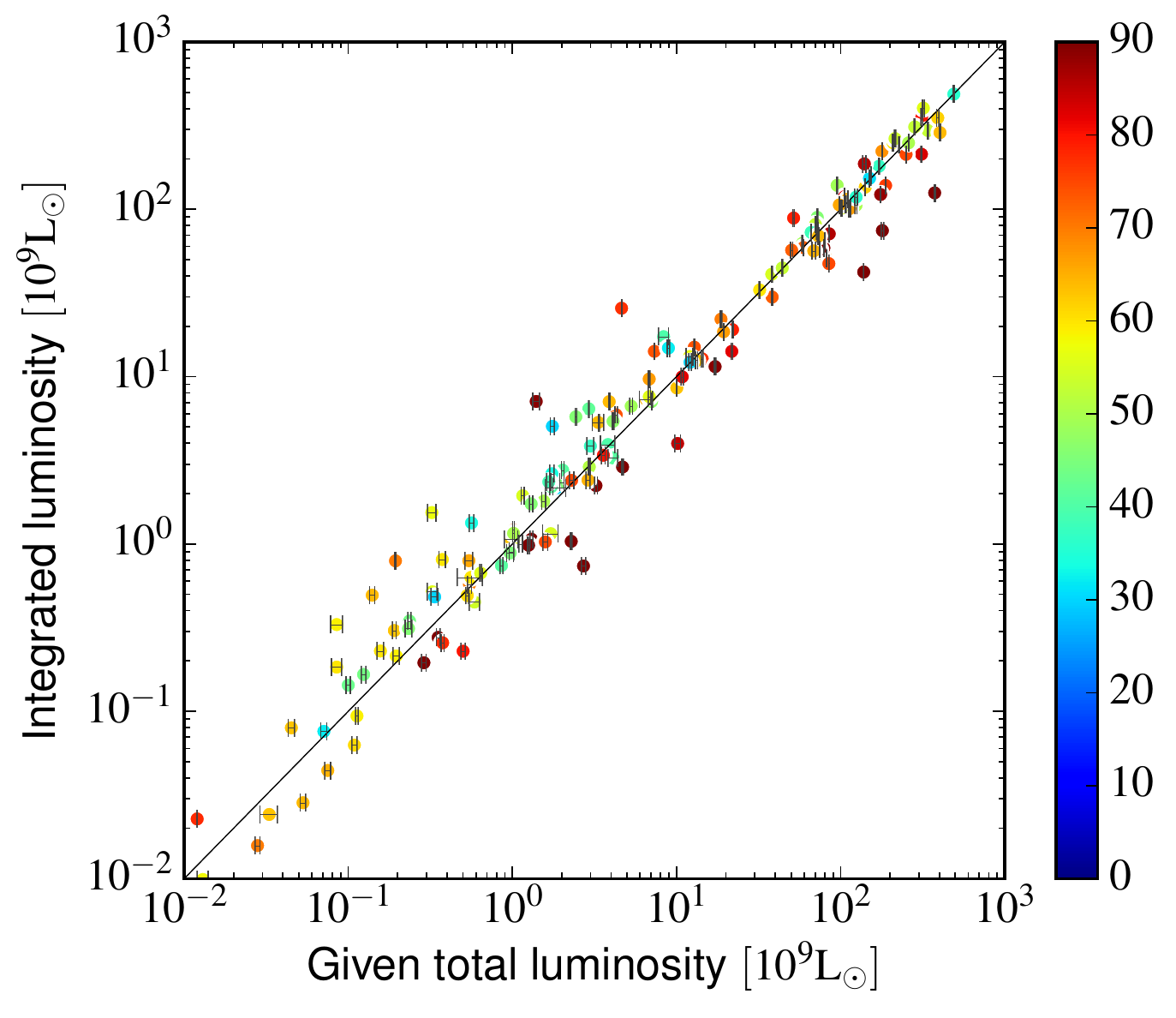}
  \caption{Comparison of given total luminosities $L_{[3.6]}$ with
    integrated luminosities $L_{\rm mod}$ for 153 galaxies.  Error
    bars show the given uncertainties, $\delta L_{[3.6]}$.  Color
    indicates the galaxy inclination angle in degrees.}\label{fig.SB36ratio}
\end{figure}

In spite of the global correlation demonstrated in Fig.\ 3 of MLS and
Fig.\ \ref{fig.histNewton}, substantial qualitative and quantitative
differences exist for many individual galaxies between the observed
rotational velocity curves and velocity curves calculated using the
RAR relation (\ref{Empiricalacc}), or Eq.\ (\ref{accelerationlaw2})
for MOG.  These differences will be due, in part, to limitations of
the underlying data, discrepancies between the data and true values
and, for MOG, the need to adjust $\alpha_\infty$, $D_0$ and $E_0$.
For example, the SPARC data includes values for total luminosity
$L_{[3.6]}$, the uncertainties $\delta L_{[3.6]}$, and stellar
luminosity profiles $\Sigma_{\rm obs}(r)$.  From the latter, we have
determined model face-on profiles
$\Sigma_{\rm mod,FO}(r)=\Sigma_{\rm disk,FO}(r)+\Sigma_{\rm
  bul,FO}(r)$, which we integrated to obtain total luminosities:
\begin{equation}
L_{\rm mod}=\int_0^{4.5\,R_{\rm d}}{\Sigma_{\rm mod,FO}(r)
  2\pi\,r\,dr}\,.
\label{eq.Lmod}
\end{equation}
Fig.\ \ref{fig.SB36ratio} compares values of $L_{[3.6]}$ to
$L_{\rm mod}$.  Differences $\Delta L=L_{\rm mod}-L_{[3.6]}$ generally
far exceed the $\pm\delta L_{[3.6]}$ error bars.  Explaining these
discrepancies does not seem feasible without information, that has not
been provided, about how the given $L_{[3.6]}$ and surface brightness
profiles were derived from the 2-d galaxy images.  Further discussion
and analysis of this issue is provided in Appendix~\ref{app.appAlt}.

\begin{figure}[t!]
  \centering \includegraphics[width=3.25in]{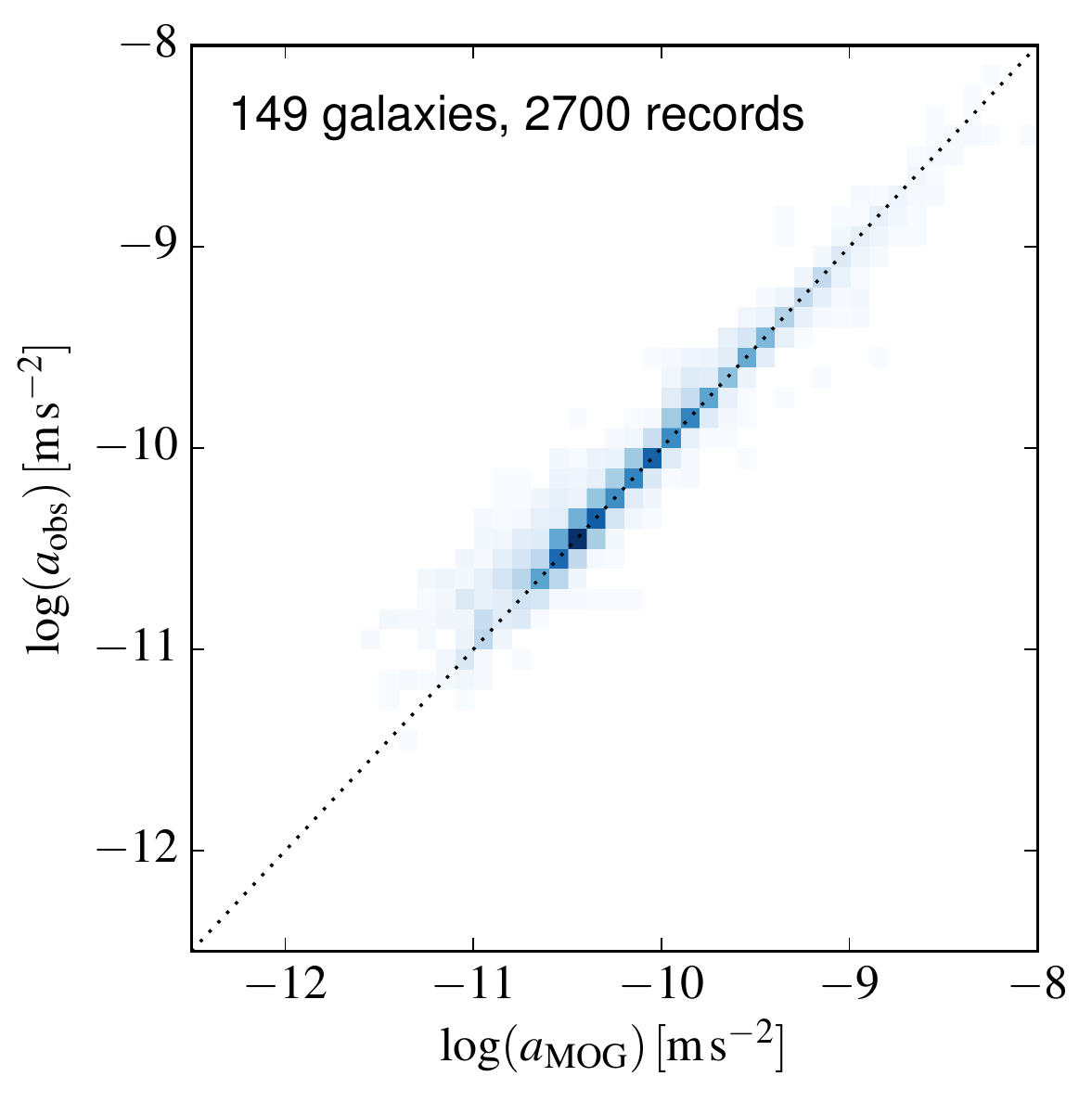}

  \includegraphics[width=3.25in]{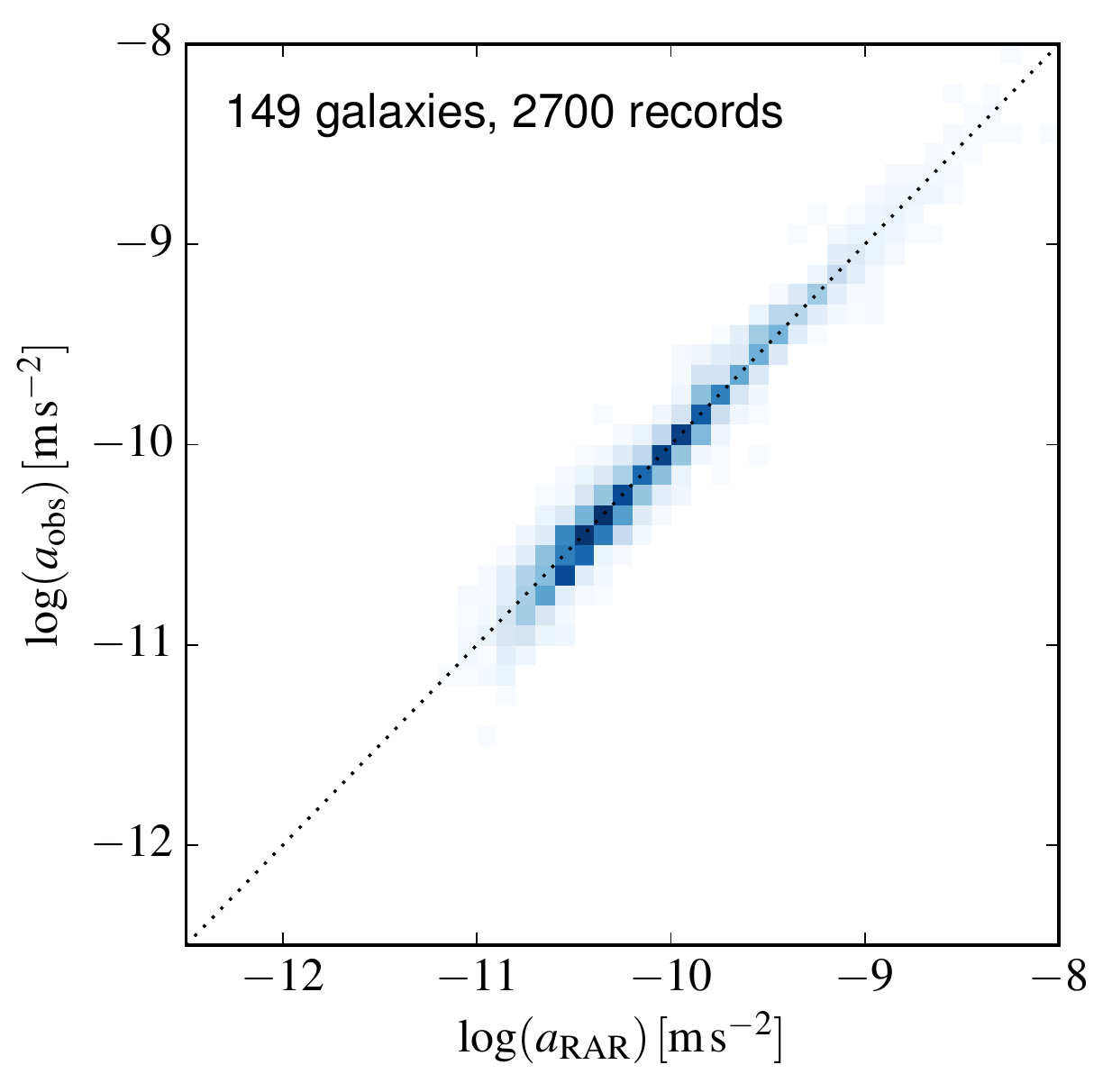}
  \caption{Top panel: Calculated $a_{\rm MOG}$ accelerations on the
    horizontal axis, similar to Fig.\ \ref{fig.histMOG}, but with
    galaxy parameters adjusted as described in the text.  The mean
    $\langle\log(a_{\rm obs}/a_{\rm MOG})\rangle = 0.05$~dex and
    residuals about the linear fit, which skew upwards at low
    acceleration, have a width $\sigma=0.13$~dex.  Bottom panel: Same
    galaxy parameters as in top panel, with calculated
    $a_{\rm Newton}$ accelerations transformed to $a_{\rm RAR}$ using
    (\ref{Empiricalacc}), with
    $a_0=(5.4\pm .3)\times 10^{-11}\,{\rm m/s^2}$\,.  The mean
    $\langle\log(a_{\rm obs}/a_{\rm RAR})\rangle = -0.006$~dex;
    residuals have a width
    $\sigma=0.11$~dex.}\label{fig.histAdjMOG_RAR}
\end{figure}

Errors in several scalar parameters $\Theta$ will affect the accuracy
of galaxy mass models and the match between observed rotational
velocities, $v_{\rm obs}$, and predicted velocities, $v_{\rm th}$,
based on a chosen theory and the modeled $\rho_\Theta(r,z)$.  Besides
the total luminosity, discussed above, these parameters include galaxy
distance, $D$, inclination, $i$, disk mass-to-light ratio,
$\Upsilon_*$, the ratio $\Upsilon_{\rm bul}/\Upsilon_*$, and total gas
mass, $M_{\rm gas} = \eta(T)\,M_{\rm HI}$.  Limiting the range to
$\pm 1$-sigma from the given values, we varied these parameters to
obtain the best fits of $v_{\rm MOG}(r)$ to $v_{\rm obs}(r)$.  In the
fits, observed velocities were weighted by
$v_{\rm obs}(r)/\delta v_{\rm obs}(r)$.  For $D$ and $i$ we used the
given uncertainties.  For luminosity uncertainty, we combined in
quadrature the given $\delta L_{[3.6]}$, $\Delta L$, and
$0.2\,L_{[3.6]}$, where the last term is a surrogate for 20\%
uncertainty of $\Upsilon_*$.  We constrained the ratio
$\Upsilon_{\rm bul}/\Upsilon_*$ to the range (1.1,1.8) and allowed for
20\% uncertainty of $M_{\rm gas}$.  The top panel of Fig.\
\ref{fig.histAdjMOG_RAR} compares $a_{\rm obs}$ to $a_{\rm MOG}$ using
galaxy parameters fitted as described above, using MOG as the theory.
The bottom panel of Fig.\ \ref{fig.histAdjMOG_RAR} compares
$a_{\rm obs}$ to $a_{\rm RAR}$, calculated using Eq.\
(\ref{Empiricalacc}) with the same galaxy parameters (adjusted to fit
MOG) and the value $a_0=(5.4\pm .3)\times 10^{-11}\,{\rm m/s^2}$ that
best fits the RAR to the adjusted galaxies.  Plots provided in the
Supplementary Material show the observed rotational velocity curves
and the calculated Newtonian, MOG and RAR predictions, with adjusted
parameters, for 153 galaxies.

\begin{figure*}[t!]
  \centering
  \hspace*{.05in}\includegraphics[height=2.6in]{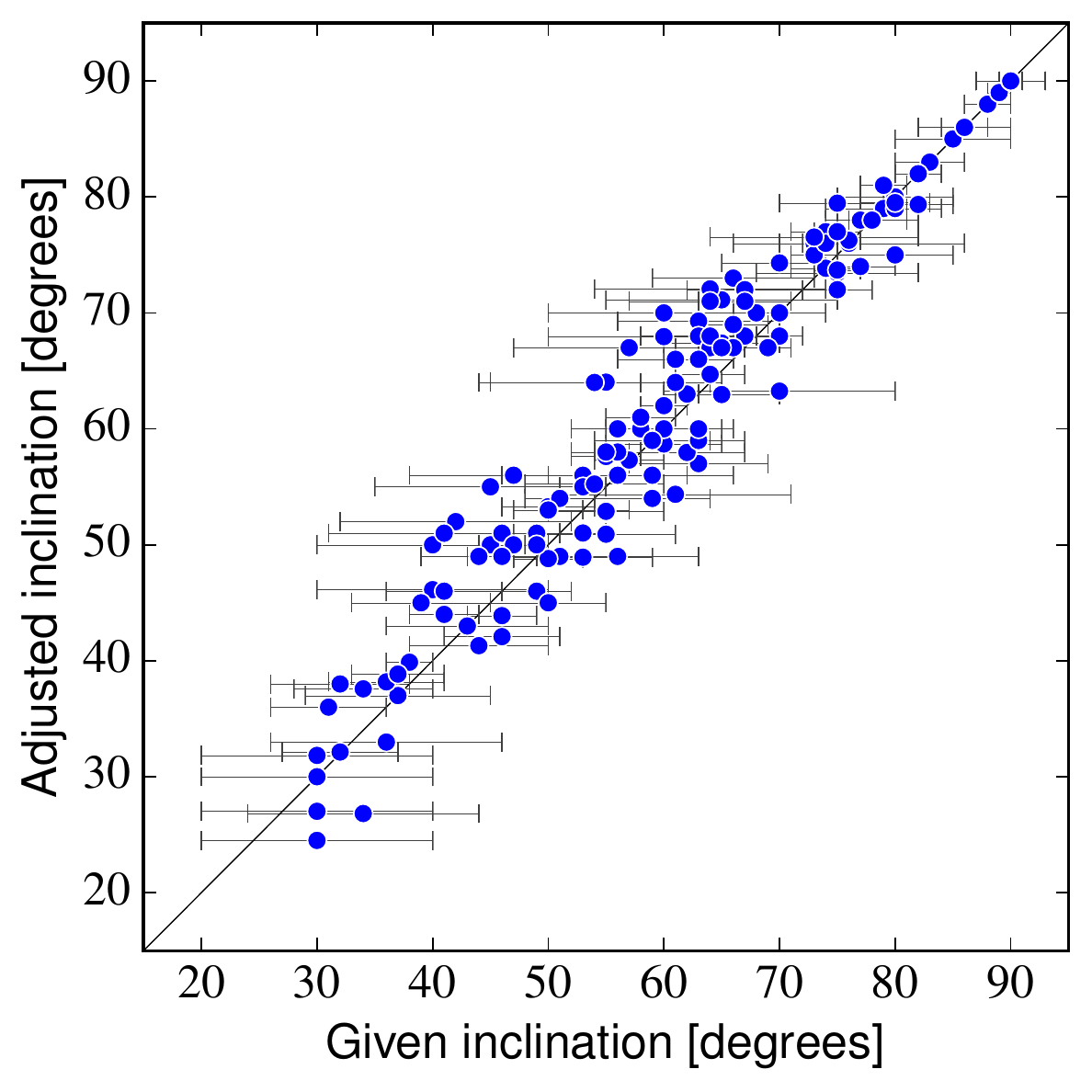}
  \hspace*{.75in}\includegraphics[height=2.6in]{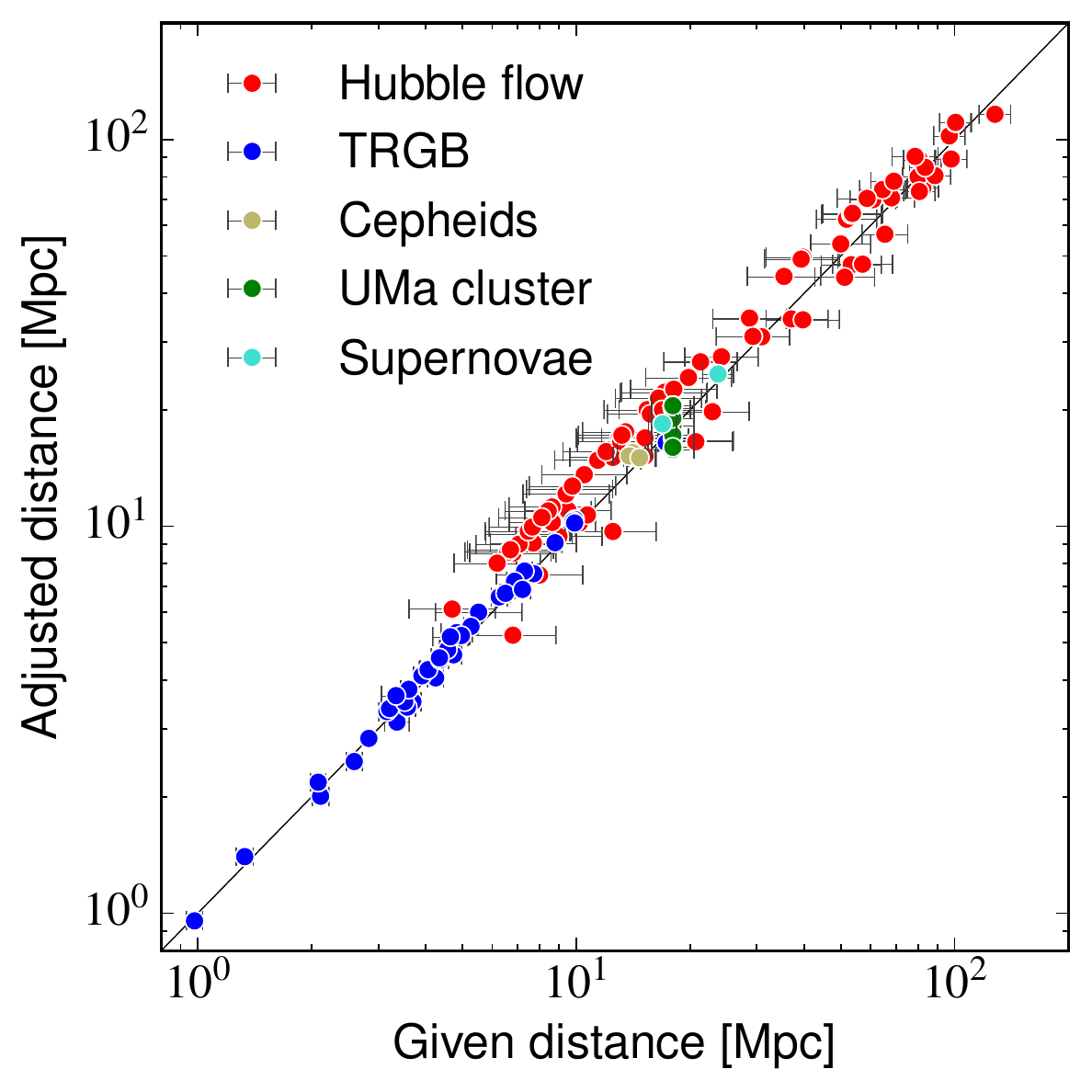}

  \includegraphics[height=2.85in]{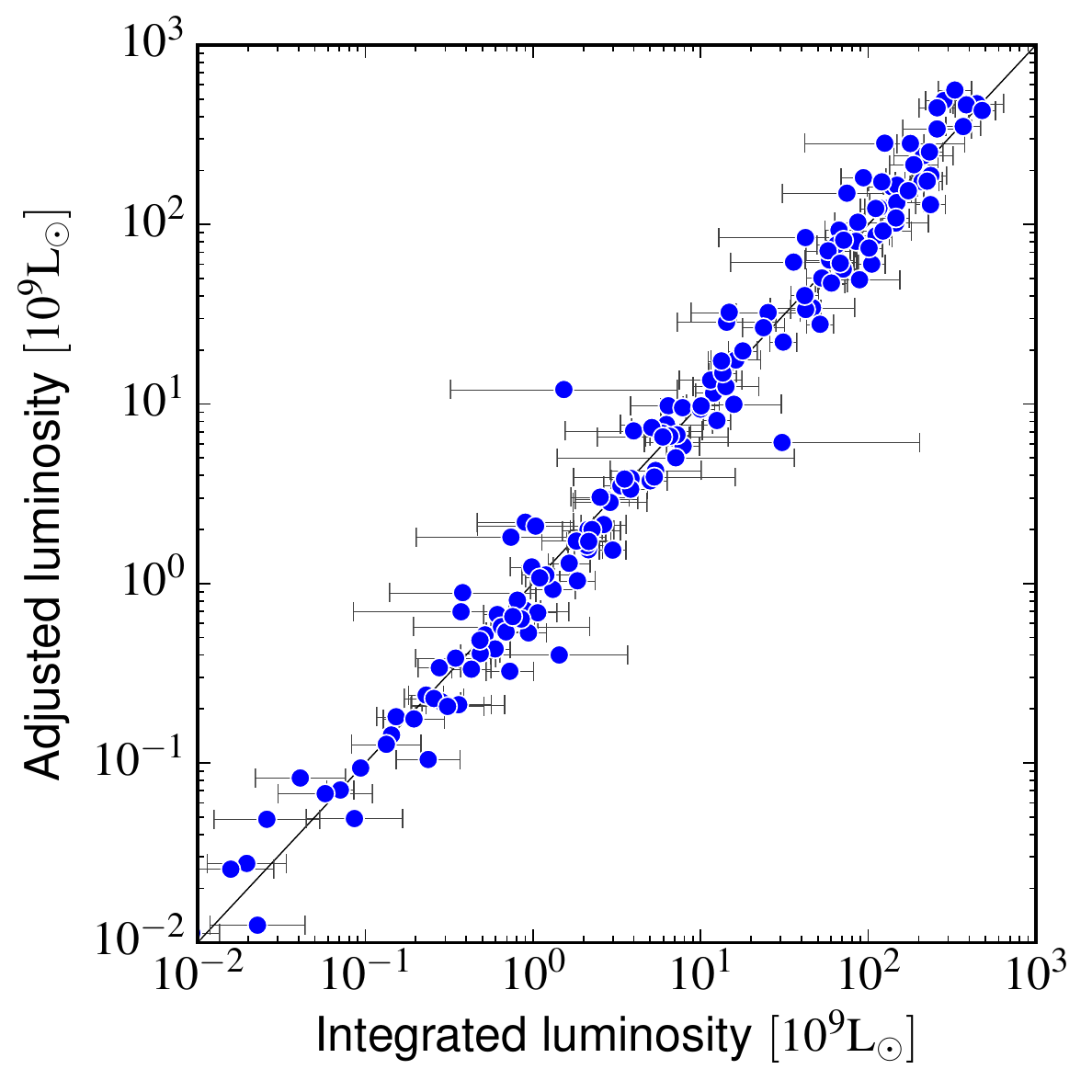}
  \hspace*{.5in}\includegraphics[height=2.85in]{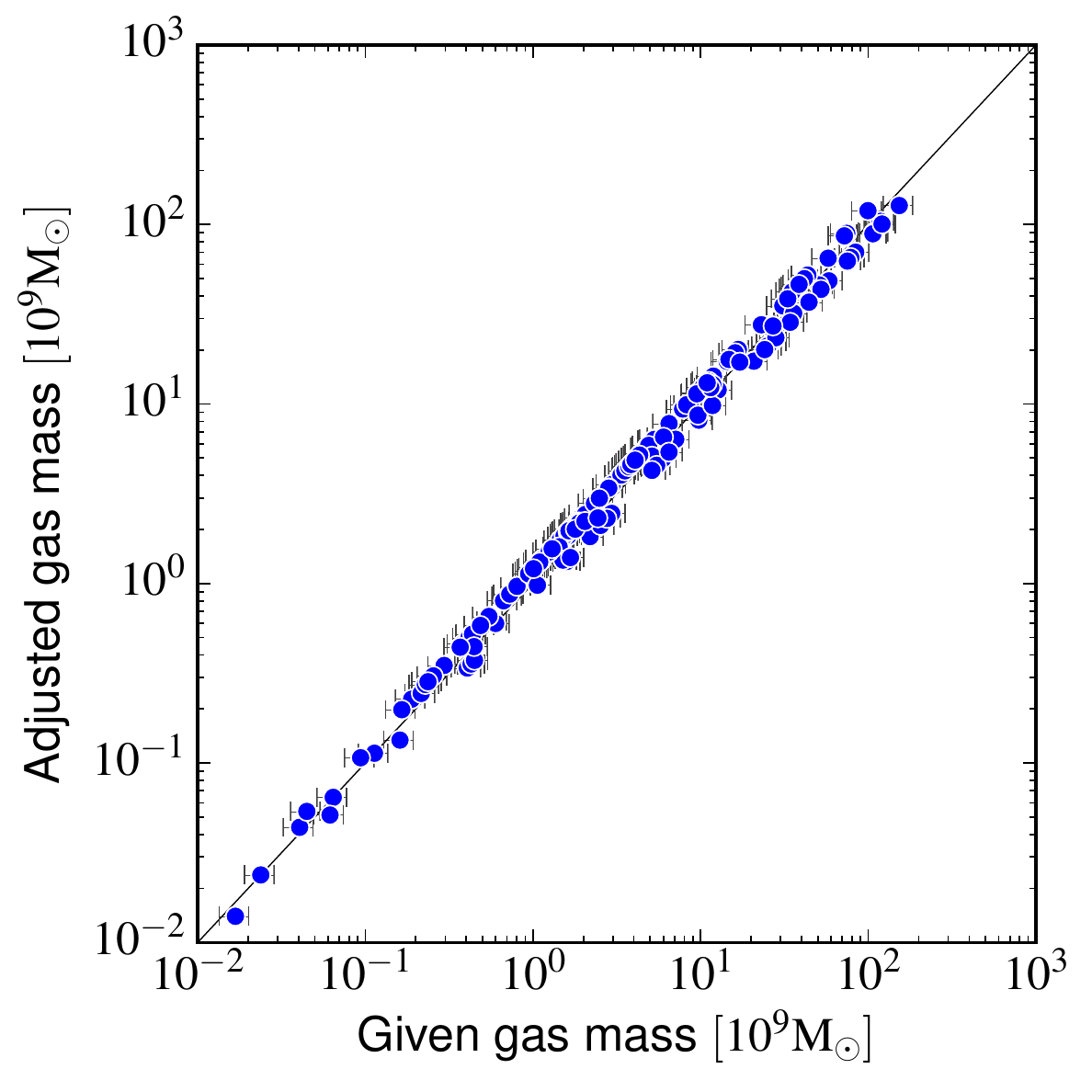}
  \caption{Comparisons of unadjusted to adjusted values of galaxy
    inclination, distance, integrated luminosity and gas mass.  For
    distances, the legend indicates the method used to determine the
    given values.  Error bars are $\pm 1$-sigma, as described in the
    text.}\label{fig.adjustments}
\end{figure*}
\begin{figure}[t!]
  \centering \includegraphics[width=2.7in]{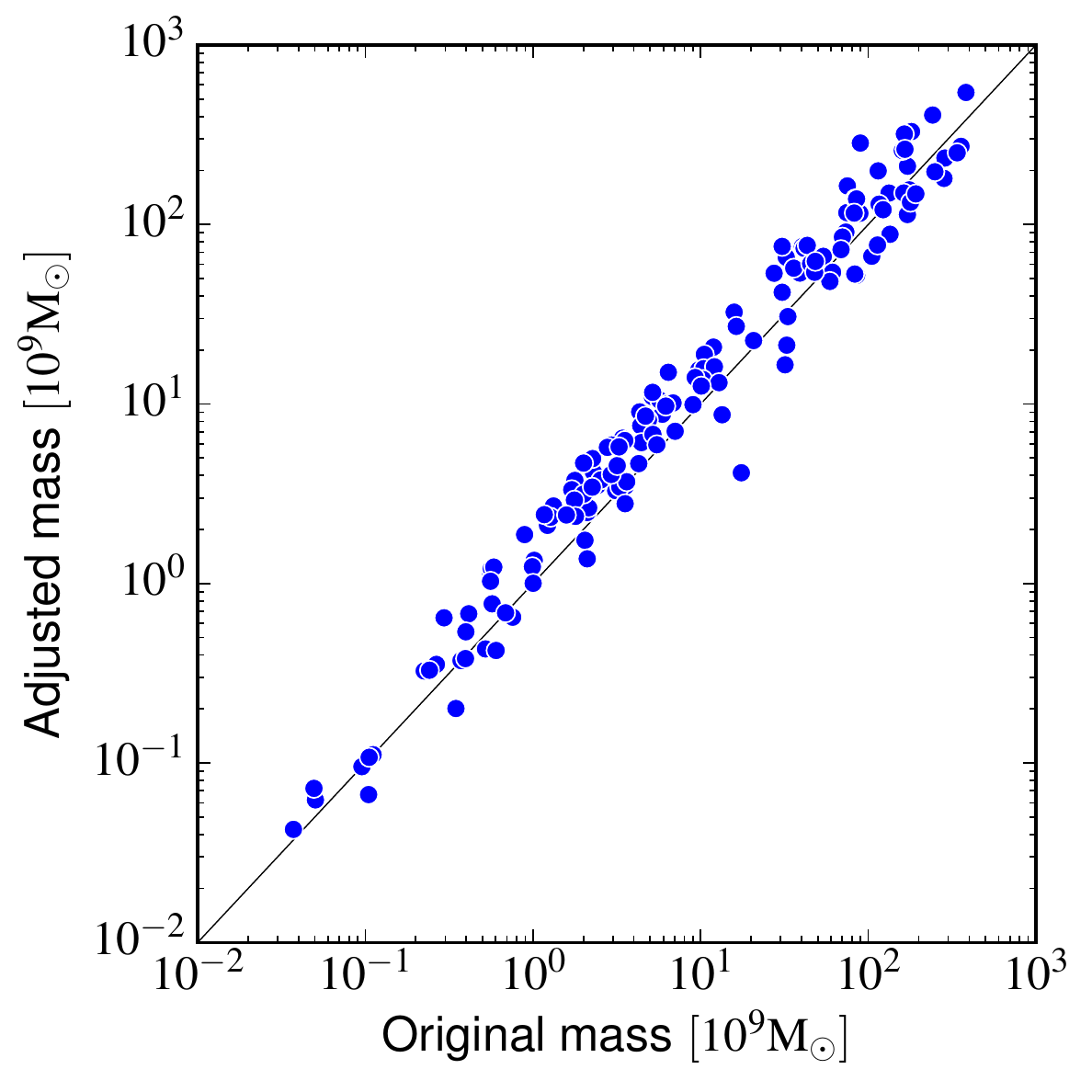}
  \caption{Comparison of original and adjusted total
    masses.}\label{fig.masses}
\end{figure}

Fig.\ \ref{fig.adjustments} shows comparisons of adjusted inclination,
distance, integrated luminosity and gas mass to the original,
unadjusted values.  (The ``given'' gas masses are
$M_{\rm gas}=\eta(T) M_{\rm HI}$.)  Fig.\ \ref{fig.masses} compares
the original and adjusted galaxy total masses.  In a great number of
cases, the adjustments were limited by our 1-sigma bounds on parameter
variations.  Nonetheless, the matches between $a_{\rm obs}$ and both
$a_{\rm MOG}$ and $a_{\rm RAR}$ in Fig.~\ref{fig.histAdjMOG_RAR} are
quite tight.  Similarity of the top and bottom panels of Fig.\
\ref{fig.histAdjMOG_RAR} indicates strong correlation of accelerations
calculated using either MOG or the RAR.

\section{Discussion}

Many significant uncertainties affected our analysis of the SPARC
galaxy data.  Uncertainty of total luminosity is highlighted by the
discrepancies between $L_{[3.6]}$ and $L_{\rm mod}$ shown in Fig.\
\ref{fig.SB36ratio}; the standard deviation of the ratios
$L_{\rm mod}/L_{[3.6]}$ for the 153 selected galaxies is 0.2 dex.
Galaxy distances determined using Hubble flow have the greatest
uncertainty, with errors of 0.1~dex or more, perhaps systematic,
expected at smaller distances due to increased significance of
peculiar velocities and local flows.  Inferred masses grow
quadratically with distance; distance errors affect $a_{\rm obs}$,
$a_{\rm Newton}$ and $a_{\rm MOG}$ differently.  The ratio
$M_{\rm gas}/M_{\rm HI}$ has large uncertainty; and the distribution
$\rho_{\rm gas}$ of all gases cannot be reliably inferred from the HI
radial profile.  Stellar [3.6] mass-to-light ratios for individual
galaxies have about 0.1~dex uncertainty \cite{Meidt}.  Errors in
inclination angles will affect estimates of both mass density and
rotational velocity.

Bulge, stellar disk and gas disk model profiles are idealizations that
ignore bars, spiral arms, disk warps, non-coplanarity of stellar and
gas disks, and galaxy interactions or other disruptions.  Such
irregularities can have $O(1)$ local effects on rotational velocities
at small radii and can result in models not reliably representing the
mass distribution.  Decomposing radial luminosity profiles into bulge
and stellar disk contributions involves subjective judgment.  Stellar
and gas disk scale heights, determined solely by formula instead of
observations, have unknown errors that affect the 3-d mass models and
calculated radial gravitational accelerations.

Rotational velocity curves were derived from HI/H$\alpha$ observations
of gas; the assumption that stellar velocities are similar to gas
velocities may not be valid if the stellar and gas disks do not have
similar thickness or if they are not coplanar.  Rotational velocities
at small radius will be uncertain because different depths along the
line-of-sight will be at different radii and angular positions.
Rotational velocities are generally more uncertain, and likely to be
underestimated, near the galactic center \cite{SofueRubin}.

Some of our modeling assumptions differ significantly from assumptions
used in MLS.  Appendix~\ref{app.appAlt} examines the effects of
switching to the MLS assumption $\mu=1.33$ and scaling of face-on
models to match $L_{[3.6]}$.  Fig.\ \ref{fig.histS133F} shows that
these changes lead to histograms very similar to those of Fig.\
\ref{fig.histAdjMOG_RAR}, indicating that these assumptions are not
material to conclusions regarding MOG.

The uncertainties listed above, including systematic errors, will
undoubtedly have contributed to the dispersion and offset from the
diagonal seen in the log-log plots of Figs.~\ref{fig.histNewton} and
\ref{fig.histMOG}.  Tuning galaxy parameters to fit a given
gravitational acceleration model, such as the RAR (\ref{Empiricalacc})
or MOG acceleration (\ref{accelerationlaw2}), will, by design, reduce
the dispersion and make results more closely match the fitted model.
Thus Fig.\ 4 of \cite{Li et al.(2018)} shows a very tight distribution
about the RAR when $\Upsilon_*$, $D$, and $i$ are optimized; and the
top panel of our Fig.~\ref{fig.histAdjMOG_RAR} shows a tight match
between $a_{\rm obs}$ and $a_{\rm MOG}$.

In \cite{Li et al.(2018)}, also optimizing $a_0$, but imposing a
narrow Gaussian prior with mean $1.2\times 10^{-10}\,{\rm m/s^2}$,
gave the histogram an even tighter distribution.  But the broad
distribution of optimal $a_0$ values obtained when using a flat prior
casts doubt on the justification of a narrow prior.  We have found,
nonetheless, that adjusting galaxy parameters to fit MOG also yields a
tight fit to the RAR, with
$a_0=(5.4\pm .3)\times 10^{-11}\,{\rm m/s^2}$.  It is shown in
Appendix~\ref{app.appAlt} that nearly the same value of $a_0$ is
obtained with modeling assumptions closer to those of MLS.

We have verified that adjusting the MOG parameters $\alpha_\infty$,
$D_0$ and $E_0$ of Eq. (\ref{eq.alpha-mu}), which are fixed for all
galaxies, can make the histogram of Fig.\ \ref{fig.histMOG} and the
adjusted mass distribution of Fig.\ \ref{fig.masses} better centred on
the diagonal.  However, we have not adjusted or optimized the MOG
parameters because we believe that the many other uncertainties and
modeling assumptions would seriously diminish the significance of the
results.  The values of galaxy parameters adjusted to fit MOG depend
on the choice of $\alpha_\infty$, $D_0$ and $E_0$, and would be
different for different MOG parameter values.

Perhaps our most surprising result is that galaxy parameters adjusted
within $\pm 1 \sigma$ bounds to fit MOG also give a very good match
between $a_{\rm obs}$ and $a_{\rm RAR}$ as shown in the bottom panel
of Fig.~\ref{fig.histAdjMOG_RAR}.  This indicates that the MOG
acceleration law (\ref{accelerationlaw2}) gives results generically
similar to the RAR functional form (\ref{Empiricalacc}).

\section{Conclusions}

Our analysis has demonstrated that, for 149 SPARC galaxies, adjusting
galaxy parameters within $\pm 1$-sigma bounds can yield MOG
predictions consistent with the given rotational velocity data.  A
0.13~dex systematic discrepancy, prior to parameter adjustments,
between observed accelerations and MOG predictions can be attributed
to uncertainties of galaxy and/or MOG parameters.  Considering the
uncertainties involved, there is no material discrepancy between MOG
and the given SPARC galaxy data.  This contrasts with the significant
discrepancy between Newtonian gravitational accelerations and observed
accelerations demonstrated in\ \cite{McGaugh}.

\section*{Acknowledgments}

We thank Viktor Toth for helpful discussions and the anonymous referee
for valuable feedback.  This research was supported in part by
Perimeter Institute for Theoretical Physics. Research at Perimeter
Institute is supported by the Government of Canada through the
Department of Innovation, Science and Economic Development Canada and
by the Province of Ontario through the Ministry of Economic
Development, Job Creation and Trade.

\appendices
\section{MOG summary}
\label{app.MOGsummary}

A detailed introduction to MOG is given in ~\cite{Moffat1}; here we
provide a quick summary.  The MOG action is given by
\begin{align}
S=S_G+S_\phi+S_S+S_M,
\end{align}
where $S_M$ is the matter action and
\begin{align}
  S_G&=\frac{1}{16\pi}\int d^4x\sqrt{-g}\left[\frac{1}{G}(R+2\Lambda)\right],\\
  S_\phi&=\int d^4x\sqrt{-g}\left[-\frac{1}{4}B^{\mu\nu}B_{\mu\nu}+\frac{1}{2}\mu^2\phi^\mu\phi_\mu\right],\\
  S_S&=\int
       d^4x\sqrt{-g}\left[\frac{1}{G^3}\left(\frac{1}{2}g^{\mu\nu}\partial_\mu
       G\partial_\nu G-V_G\right)\right.\nonumber\\
  & ~ \left.\hspace*{.8in}+\frac{1}{\mu^2G}\left(\frac{1}{2}g^{\alpha\beta}\partial_\alpha\mu\partial_\beta\mu-V_\mu\right)\right],\label{Eq:SS}
\end{align}
where $B_{\mu\nu}=\partial_\mu\phi_\nu-\partial_\nu\phi_\mu$ and $V_G$
and $V_\mu$ are potentials.  Note that we choose units such that
$c=1$, and use the metric signature $[+,-,-,-]$. The Ricci tensor is
\begin{equation}
  R_{\mu\nu}=\partial_\lambda\Gamma^\lambda_{\mu\nu}-\partial_\nu\Gamma^\lambda_{\mu\lambda}
  +\Gamma^\lambda_{\mu\nu}\Gamma^\sigma_{\lambda\sigma}-\Gamma^\sigma_{\mu\lambda}\Gamma^\lambda_{\nu\sigma}.
\end{equation}

The matter stress-energy tensor is obtained by varying the matter
action $S_M$ with respect to the metric:
\begin{align}
  T^{\mu\nu}_M=-2(-g)^{-1/2}\delta S_M/\delta g_{\mu\nu}\,.
\end{align}
Varying $S_\phi+S_S$ with respect to the metric yields
\begin{align}
  T^{\mu\nu}_{\rm MOG}=-2(-g)^{-1/2}[\delta S_\phi/\delta
  g_{\mu\nu}+\delta S_S/\delta g_{\mu\nu}]\,.
\end{align}
Combining these gives the total stress-energy tensor
\begin{align}
  T^{\mu\nu}=T^{\mu\nu}_{\rm M}+T^{\mu\nu}_{\rm MOG}\,.
\end{align}

The MOG field equations are given by
\begin{align}
  G_{\mu\nu}-\Lambda g_{\mu\nu}+Q_{\mu\nu}=8\pi GT_{\mu\nu}\,,
  \label{eq:MOGE}
\end{align}
\begin{equation}
  \frac{1}{\sqrt{-g}}\partial_\mu\biggl(\sqrt{-g}B^{\mu\nu}\biggr)+\mu^2\phi^\nu=-J^\nu,
  \label{eq:phi}
\end{equation}
\begin{equation}
  \partial_\sigma B_{\mu\nu}+\partial_\mu B_{\nu\sigma}+\partial_\nu B_{\sigma\mu}=0.
\end{equation}
Here, $G_{\mu\nu}=R_{\mu\nu}-\frac{1}{2} g_{\mu\nu} R$ is the Einstein
tensor and
\begin{align}
  Q_{\mu\nu}=\frac{2}{G^2}(\partial^\alpha G \partial_\alpha G\,g_{\mu\nu}
  - \partial_\mu G\partial_\nu G) - \frac{1}{G}(\Box G\,g_{\mu\nu}
  - \nabla_\mu\partial_\nu G)
  \label{eq:Q}
\end{align}
is a term resulting from the the presence of second derivatives of
$g_{\mu\nu}$ in $R$ in $S_G$. The current $J^\nu$, introduced in
(\ref{eq:phi}), is discussed below.

Combining the Bianchi identities, $\nabla_\nu G^{\mu\nu}=0$, with the
field equations (\ref{eq:MOGE}) yields the conservation law
\begin{align}
  \nabla_\nu T^{\mu\nu}+\frac{1}{G}\partial_\nu G\,T^{\mu\nu} -
  \frac{1}{8\pi G}\nabla_\nu Q^{\mu\nu}=0 \,.
  \label{eq:Conservation}
\end{align}

It is a key premise of MOG that all baryonic matter possesses, in
proportion to its mass $M$, positive gravitational charge:
$Q_g=\kappa\,M$.  This charge serves as the source of the vector field
$\phi^\mu$.  Moreover, $\kappa=\sqrt{G-G_N}=\sqrt{\alpha\,G_N}$, where
$G_N$ is Newton's gravitational constant and
$\alpha=(G-G_N)/G_N\ge 0$.  Variation of $S_M$ with respect to the
vector field $\phi^\mu$, then yields the MOG 4-current
$J_\mu=-(-g)^{-1/2}\delta S_M/\delta \phi^\mu$.

For the case of a perfect fluid:
\begin{equation}
\label{energymom}
T^{\mu\nu}_M=(\rho+p)u^\mu u^\nu-pg^{\mu\nu},
\end{equation}
where $\rho$ and $p$ are the matter density and pressure,
respectively, and $u^\mu$ is the 4-velocity of an element of the
fluid.  We obtain from (\ref{energymom}) and $u^\mu u_\mu=1$ the
4-current:
\begin{equation}
J_\mu=\kappa T_{M\mu\nu}u^\nu=\kappa\rho u_\mu.
\end{equation}
It is shown in~\cite{Roshan} that, with the assumption
$\nabla_\mu J^\mu=0$, (\ref{eq:Conservation}) reduces to
\begin{equation}
\nabla_\nu T_M^{\mu\nu}=B_\nu^{~\mu} J^\nu\,.\label{eq:Mcons}
\end{equation}
It should be stated that in MOG early universe cosmology, using FRW,
we do not assume $\nabla_\mu J^\mu=0$ \cite{MoffatToth3,MoffatGreenToth,McGaugh}.

The original formulation of MOG, presented above, also has field
equations for $G$ (or $\alpha$) and $\mu$:
\begin{equation}
\Box G=K,\quad\Box\mu=L,
\end{equation}
where $\Box=\nabla^\mu\nabla_\mu$, $K=K(G,\mu,\phi_\mu)$ and
$L=L(G,\mu,\phi_\mu)$.  In the present work, and other recent studies,
$\alpha$ and $\mu$ are treated as parameters that are constant for
each galaxy, but whose values run with the galaxy mass.  With this
simplification: $S_S=0$, $Q_{\mu\nu}=0$,
$T^{\mu\nu}_{\rm MOG}=-2(-g)^{-1/2}\delta S_\phi/\delta g_{\mu\nu}$
and $\nabla_\nu T^{\mu\nu}=0$.

\section{Investigation of modeling differences}
\label{app.appAlt}
\renewcommand{\theequation}{\Alph{section}-\arabic{equation}}
\setcounter{equation}{0}
\setcounter{figure}{0}

\begin{figure*}[t!]
  \centering \includegraphics[width=3.25in]{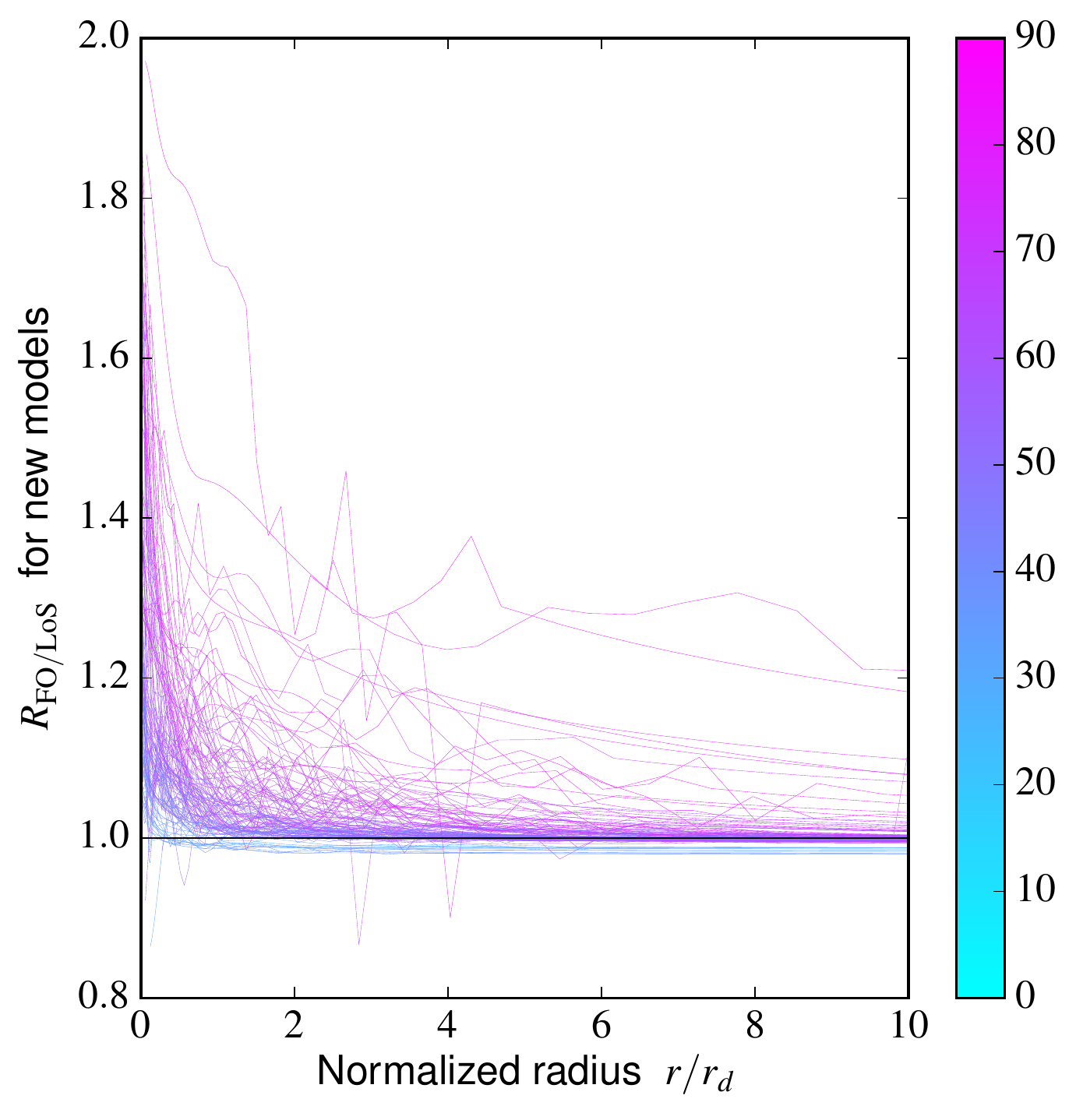}\hspace*{.25in}
  \includegraphics[width=3.25in]{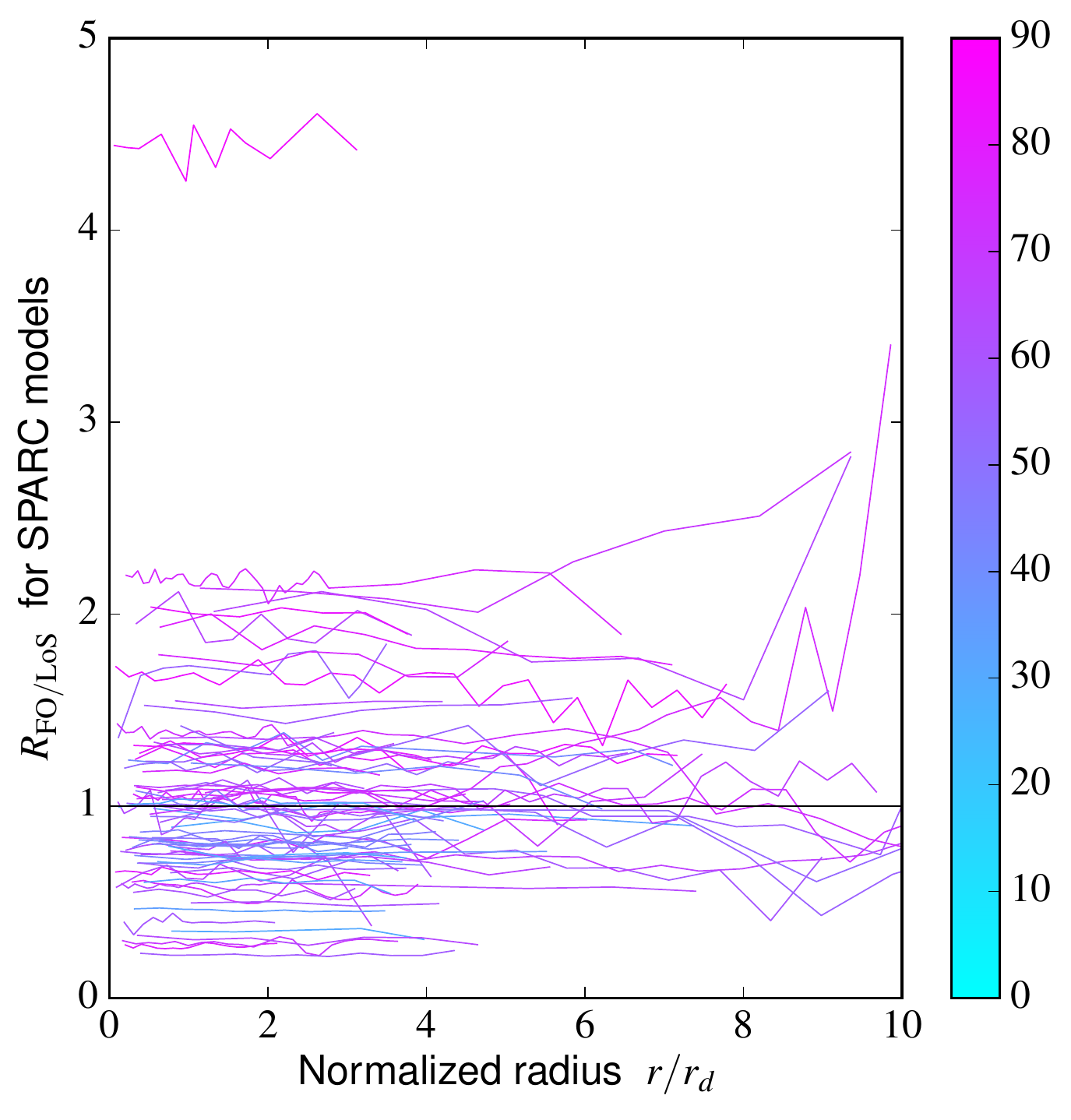}
  \caption{Left panel: Curves showing $R_{\rm FO/LoS}$ for the disks
    of 135 galaxies with $i\le 85^\circ$, based on the new galaxy
    models developed as described in Section~\ref{sec:SPARC-MOG}.
    Colour indicates the galaxy inclination in degrees.  Right panel:
    Similar to left panel, but using the $\Sigma_{\rm disk,FO}(r)$
    values provided with the SPARC data, and showing only 79 galaxies
    with $i\le 85^\circ$ and for which no bulge was identified in
    either the present or the SPARC analysis.}\label{fig.R-LoS}
\end{figure*}

Our development of galaxy mass models $\rho_{\rm bar}(r,z)$, based on
the SPARC data, was described in Section~\ref{sec:SPARC-MOG}.  Some
assumptions made in the derivation of these new models differ from
apparent and explicit assumptions in MLS.  Instead of explicit mass
models, the provided SPARC data includes face-on surface brightness
profiles $\Sigma_{\rm disk,FO}(r)$, for radii $r$ at which rotational
velocities were also given.  How the SPARC $\Sigma_{\rm disk,FO}(r)$
values were obtained was not specified.  We examine here the
relationships between line-of-sight and face-on surface brightness
profiles from the SPARC data and between equivalent profiles for our
new models.  We also examine the effects of changing one or both of
two significant assumptions made in the derivation of our new models
to match those of MLS.

For an axially-symmetric thin disk the line-of-sight and face-on
surface brightness should generally satisfy:%
\begin{equation}%
  R_{\rm FO/LoS}\equiv\frac{\Sigma_{\rm disk,FO}}
  {\cos(i)\Sigma_{\rm disk,LoS}}\simeq 1\,.\label{eq.R-FO-LoS}%
\end{equation}%
With finite scale height, $z_d$, the approximation will fail at radii
$r\lesssim z_{\rm d}/\cos(i)$\, because Eq.~(\ref{eq.R-FO-LoS}) does
not consider the exponential disk profile.  For nearly edge-on
inclinations, $R_{\rm FO/LoS}$ will not be useful because $1/\cos(i)$
diverges.

\begin{figure}[t!]
  \centering \includegraphics[width=3.25in]{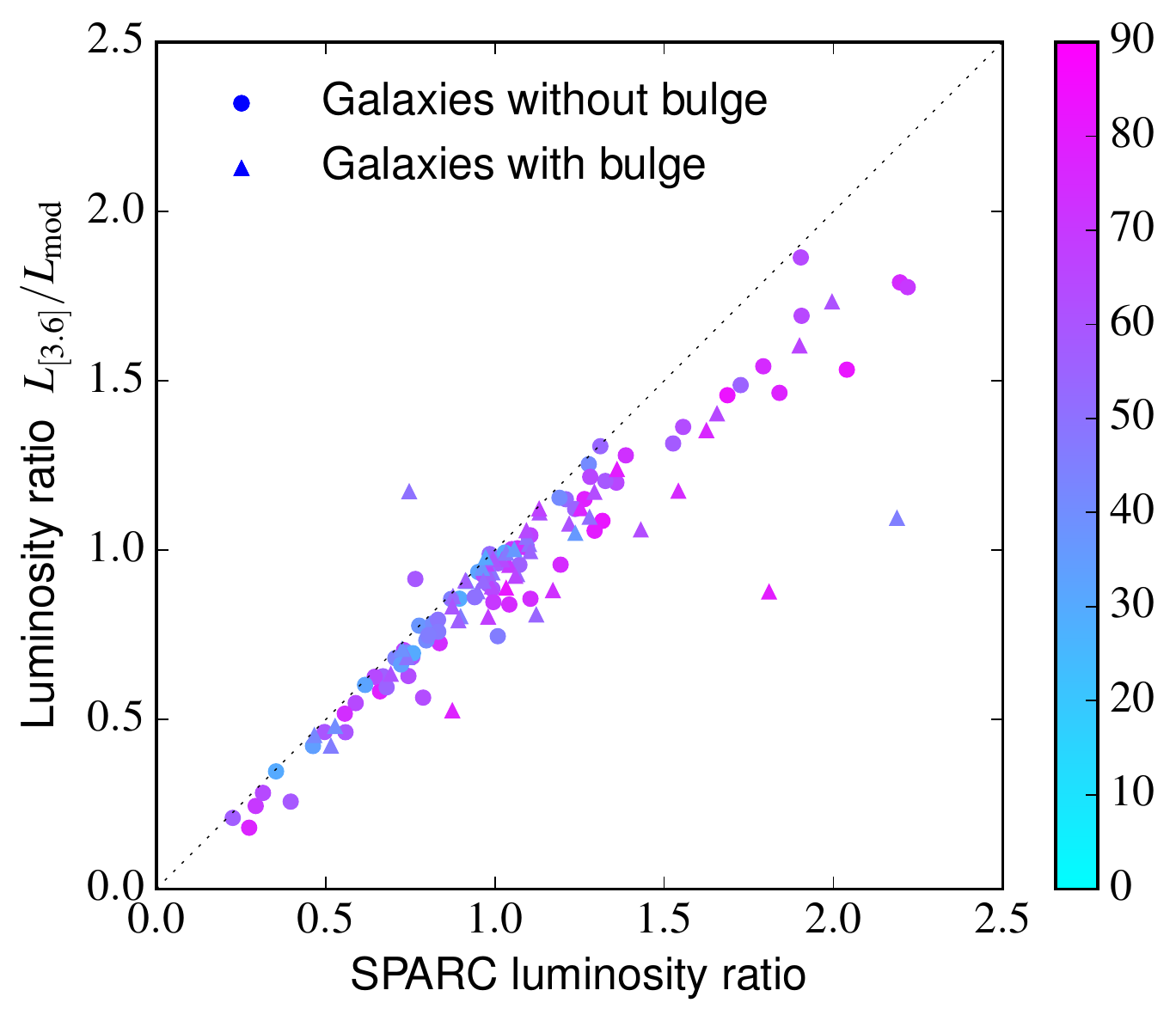}
  \caption{Comparison of the luminosity ratios
    $R_{[3.6]}=L_{[3.6]}/L_{\rm mod}$ for the new galaxy models to
    SPARC luminosity ratios, defined as the median of SPARC
    $R_{\rm FO/LoS}$ values (as in the right panel of
    Fig.~\ref{fig.R-LoS}), with extreme points excluded.  Of the 135
    galaxies with $i\le 85^\circ$, bulges were identified in 56
    galaxies in our new models and/or the SPARC
    models.}\label{fig.lumRatioComp}
\end{figure}

Fig.~\ref{fig.R-LoS} shows plots of $R_{\rm FO/LoS}$ versus normalized
radius $r/r_d$ for our new galaxy models (left panel) and as derived
directly from the SPARC data (right panel).  The plots for the new
galaxy models have the expected characteristics; the trend of low
inclination galaxies to have $R_{\rm FO/LoS}$ slightly below 1 can be
attributed to approximation in the iterative determination of
$\Sigma_{\rm mod,FO}$.  The plots using $\Sigma_{\rm disk,FO}$ and
$\Sigma_{\rm disk,LoS}$ from the SPARC data are not consistent with
expectations.  Some of the fluctuations can be attributed to limited
significant digits of the provided tabular data.  Ignoring the
quantization noise due to rounding / truncation of the data,
especially at large radii, it appears as though the face-on data has
been obtained by an almost constant, often substantial, scaling of the
line-of-sight data.

\begin{figure*}[t!]
  \centering \includegraphics[width=3.25in]{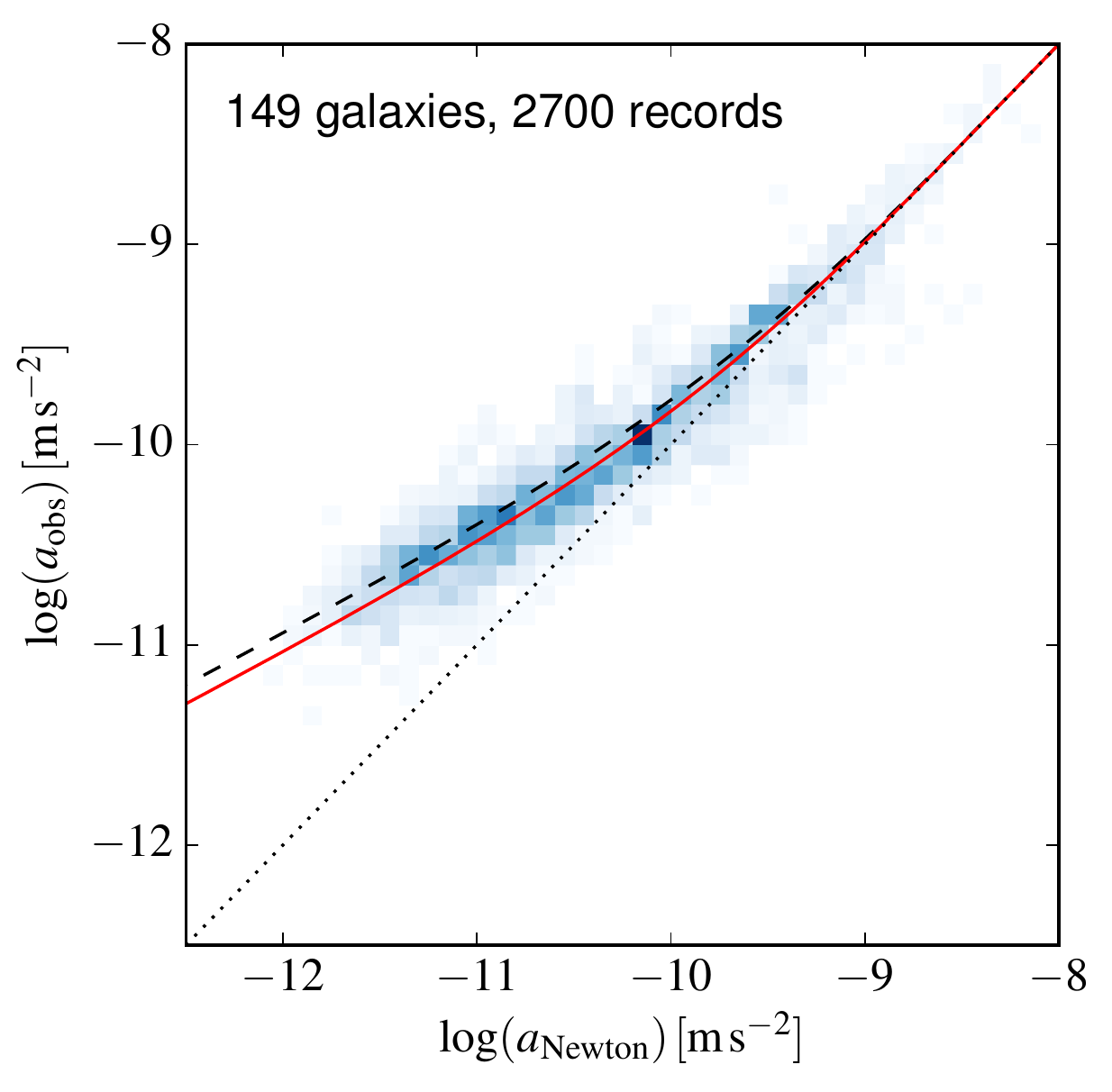}\hspace*{.25in}
  \includegraphics[width=3.25in]{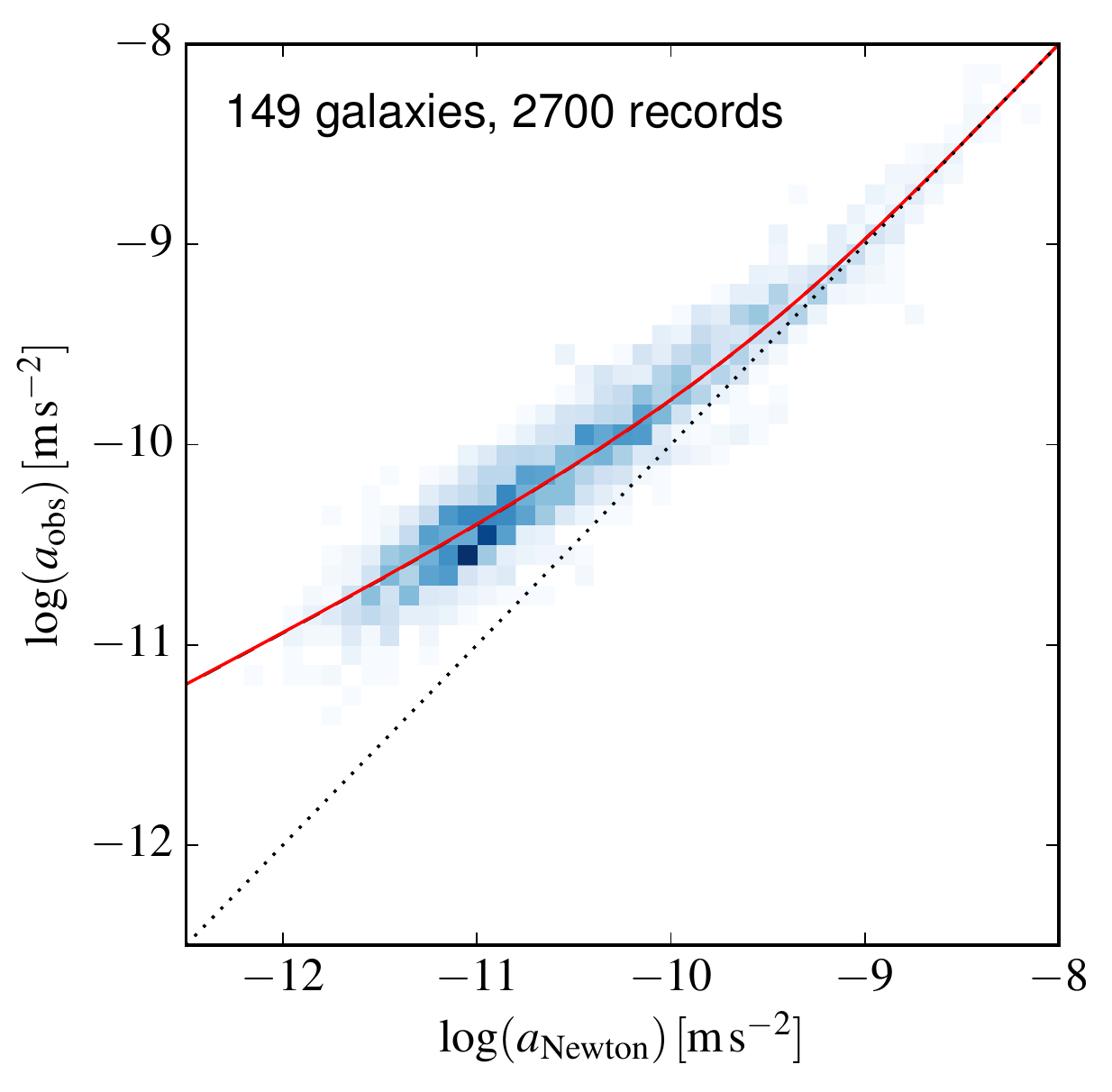}
  \caption{Left panel: Relation of $a_\mathrm{obs}$ to
    $a_\mathrm{Newton}$ when $\Sigma_\mathrm{mod,FO}$, for each
    galaxy, has been scaled to make $R_{[3.6]}=1$.  The dashed black
    curve is the RAR (\ref{Empiricalacc}) with
    $a_0=1.2\times 10^{-10}\,{\rm m/s^2}$.  The solid red curve shows
    the RAR with best fit
    $a_0=(7.6\pm.6)\times 10^{-11}\,{\rm m/s^2}$.  Residuals relative
    to the latter curve have a width $\sigma=0.16$~dex.  Right panel:
    Similar to left panel, but with galaxy mass models that, instead
    of scaling $\Sigma_\mathrm{mod,FO}$, have
    $\mu=M_{\rm gas}/M_{\rm HI}=1.33$.  The red curve uses the best
    fit $a_0=(1.2\pm.1)\times 10^{-10}\,{\rm m/s^2}$.  Residuals
    relative to the red curve have a width
    $\sigma=0.14$~dex.}\label{fig.histS_133}
\end{figure*}
\begin{figure*}[t!]
  \centering \includegraphics[width=3.25in]{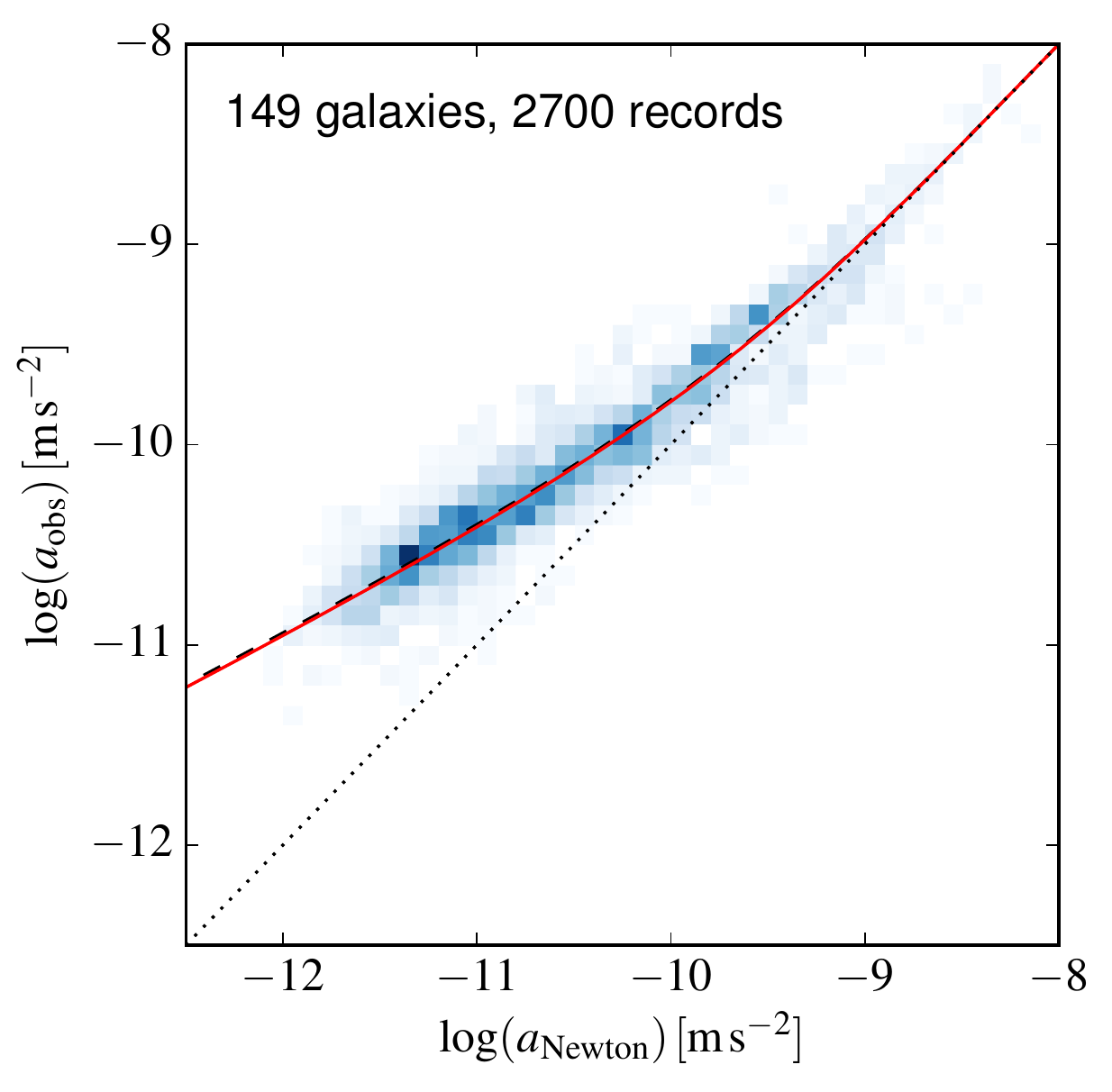}\hspace*{.25in}
  \includegraphics[width=3.25in]{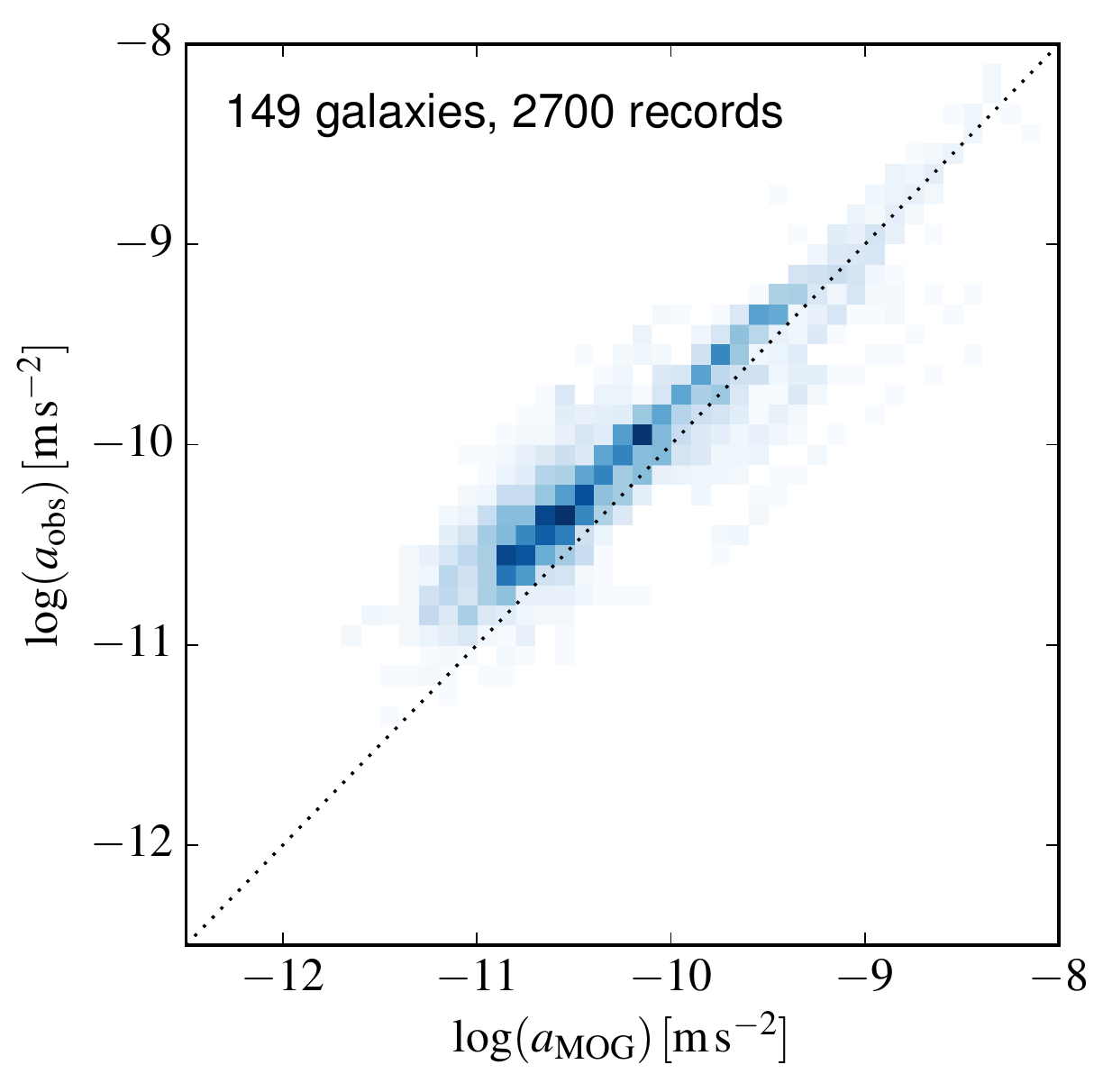}
  \caption{Left panel: Relation of $a_\mathrm{obs}$ to
    $a_\mathrm{Newton}$ with $\Sigma_\mathrm{mod,FO}$ scaled to make
    $R_{[3.6]}=1$ and with $\mu=M_{\rm gas}/M_{\rm HI}=1.33$.  The
    dashed black curve is the RAR (\ref{Empiricalacc}) with
    $a_0=1.2\times 10^{-10}\,{\rm m/s^2}$.  The solid red curve shows
    the RAR with best fit
    $a_0=(1.1\pm.1)\times 10^{-10}\,{\rm m/s^2}$.  Residuals relative
    to the red curve have a width $\sigma=0.15$~dex.  Right panel:
    Relation of $a_\mathrm{obs}$ to $a_\mathrm{MOG}$ for the same mass
    models as the left panel.  The mean
    $\langle \log(a_{\rm obs}/a_{\rm MOG})\rangle =0.18$~dex, and the
    residuals have width $\sigma=0.25$~dex.}\label{fig.histS133}
\end{figure*}

Fig.~\ref{fig.lumRatioComp} compares the ratio%
\begin{equation}%
  R_{[3.6]}=L_{[3.6]}/L_\mathrm{mod}%
\end{equation}%
for the new models of this work to SPARC luminosity ratios, defined as
the median value of $R_{\rm FO/LoS}$ for each galaxy (as in the right
panel of Fig.~\ref{fig.R-LoS}), that measure the anomalous scaling of
$\Sigma_{\rm disk,FO}(r)$ from the SPARC data beyond the factor of
$\cos(i)$.  Approximate equality of the $R_{[3.6]}$ and SPARC
luminosity ratios suggests that the SPARC $\Sigma_{\rm disk,FO}$ data
were obtained by simply scaling $\Sigma_{\rm disk,LoS}$ to match the
total luminosity, without considering the relation to inclination
given by Eq.~(\ref{eq.R-FO-LoS}).

For our primary analysis, the given surface brightness profiles
$\Sigma_{\rm obs}$ were used as the line-of-sight profiles of axially
symmetric galaxies characterized as described in Section
\ref{sec:SPARC-MOG}.  However, as shown in Figures\
\ref{fig.SB36ratio} and \ref{fig.lumRatioComp}, for many galaxies the
total luminosity $L_\mathrm{mod}$ obtained from Eq.\ (\ref{eq.Lmod})
differs significantly from the given $L_{[3.6]}$.  Contributions to
the discrepancy between the given and calculated total luminosities
will include: inability of profiles, $\Sigma_{\rm obs}$, along the
major axis, to fully represent the irregular 2-d images; and
deviations of our (and the SPARC collaboration's) necessarily simple
galaxy models from the true, irregular mass distributions.

Scaling $\Sigma_{\rm mod,FO}$ in our mass models to make $R_{[3.6]}=1$
is straightforward; the results before any adjustment of galaxy
parameters are shown in the left panel of Fig.\ \ref{fig.histS_133}.

We used Eq.\ (\ref{eq.eta}) to determine the $M_{\rm gas}/M_{\rm HI}$
ratio $\eta$, whereas MLS assumed $\eta=1.33$ for all galaxies.  The
result of repeating our calculations of $a_{\rm Newton}$ with
$\eta=1.33$ is shown in the right panel of Fig.\ \ref{fig.histS_133}.

The effects of these revised assumptions can be seen by comparing the
histograms of Fig.\ \ref{fig.histS_133} with
Fig. \ref{fig.histNewton}.  Scaling $\Sigma_\mathrm{mod,FO}$ to make
$R_{[3.6]}=1$ resulted in slightly lower $a_0$, but comparable
residuals.  Setting $\eta=1.33$ resulted in $a_0$ that matches the MLS
value, with modestly smaller residuals than Fig. \ref{fig.histNewton}.
Our different assumption regarding gas mass, which can only lead to a
larger $M_{\rm gas}$, seems to be the main reason we found lower $a_0$
values than MLS.

The combined effect of scaling $\Sigma_\mathrm{mod,FO}$ to make
\mbox{$R_{[3.6]}=1$} and setting $\eta=1.33$ is shown in the left
panel of Fig.\ \ref{fig.histS133}.  The resulting best fit $a_0$ is
close to the MLS value; the residuals are slightly larger than in our
Fig.\ \ref{fig.histNewton}. The right panel shows MOG predictions for
these revised mass models, and can be compared with Fig.\
\ref{fig.histMOG}.

\begin{figure*}[t!]
  \centering \includegraphics[width=3.25in]{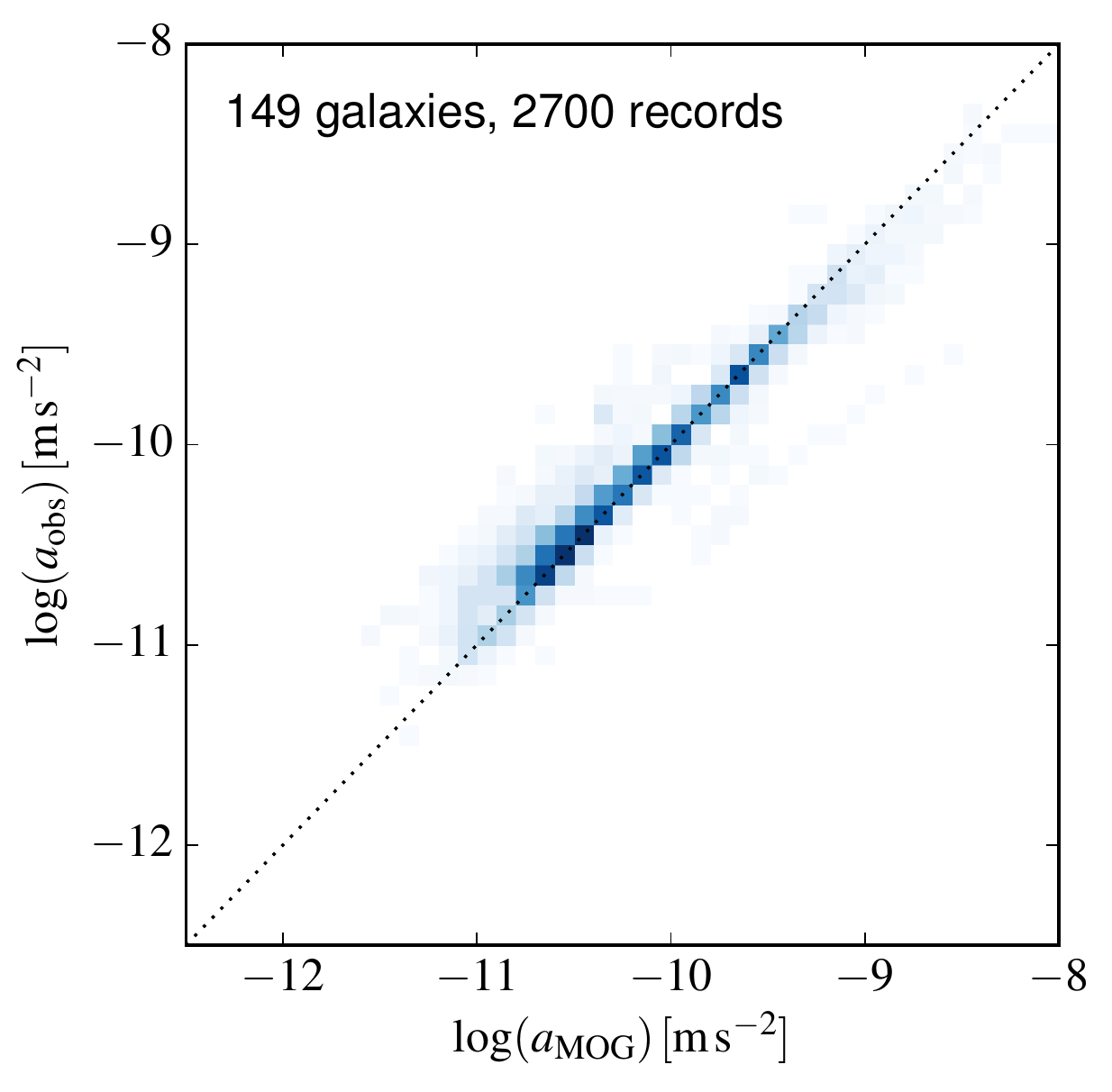}\hspace*{.25in}
  \includegraphics[width=3.25in]{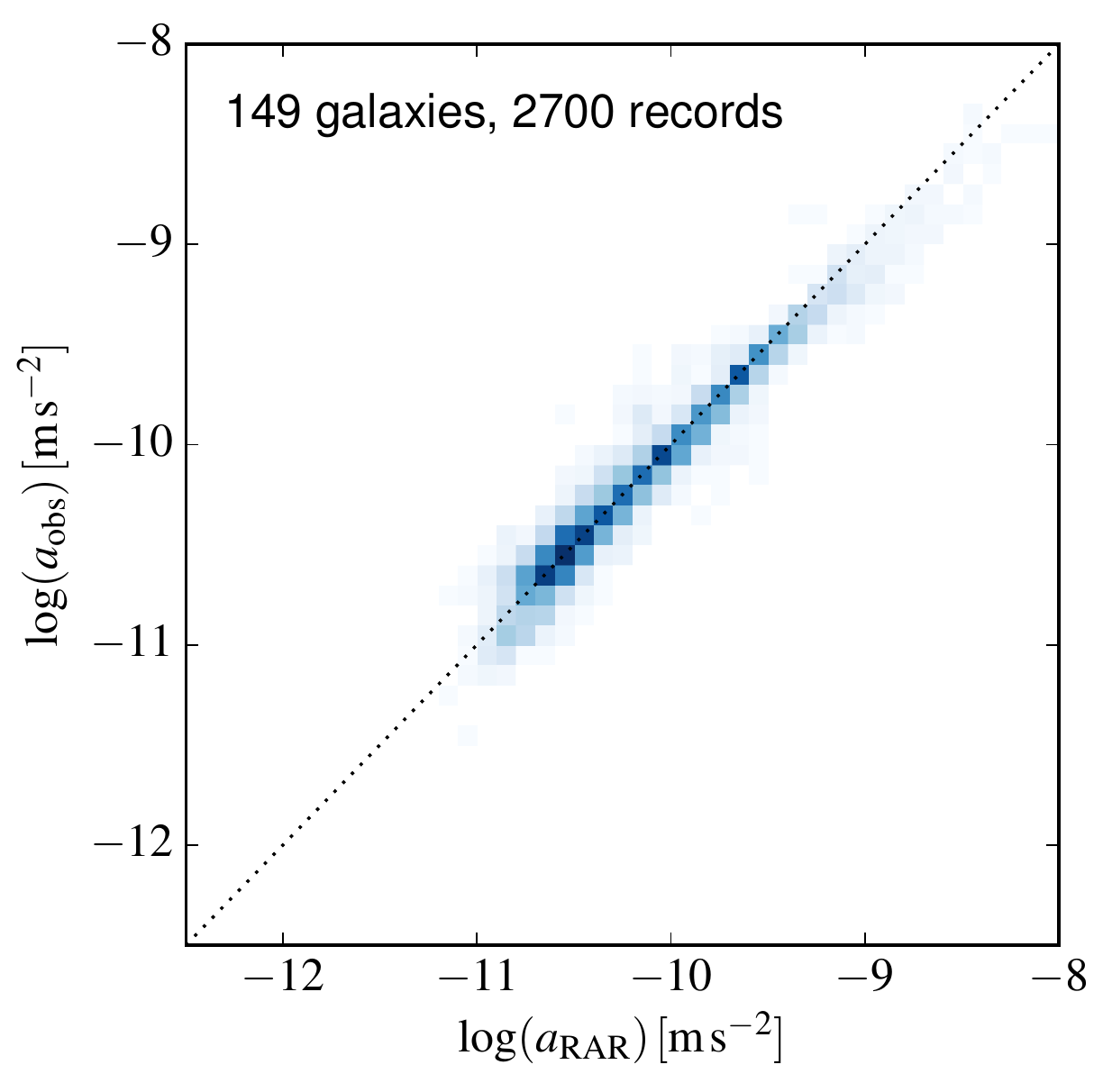}
  \caption{Left panel: Relation of $a_\mathrm{obs}$ to
    $a_\mathrm{MOG}$ for the same mass models as in
    Fig.\ref{fig.histS133} ($R_{[3.6]}=1$, $\mu=1.33$) and with galaxy
    parameters adjusted using the same process as for Fig.\
    \ref{fig.histAdjMOG_RAR}.  The mean
    $\langle\log(a_{\rm obs}/a_{\rm MOG})\rangle = 0.06$~dex;
    residuals have a width $\sigma=0.16$~dex.  Right panel: Same
    galaxy parameters as in top panel, with calculated
    $a_{\rm Newton}$ accelerations transformed to $a_{\rm RAR}$ using
    (\ref{Empiricalacc}), with
    $a_0=(5.3\pm .5)\times 10^{-11}\,{\rm m/s^2}$\,.  The mean
    $\langle\log(a_{\rm obs}/a_{\rm RAR})\rangle = 0.005$~dex;
    residuals have a width $\sigma=0.11$~dex.}\label{fig.histS133F}
\end{figure*}

Using mass models with $R_{[3.6]}=1$ and $\eta=1.33$, and adjusting
galaxy parameters using the same algorithm and bounds as used to
produce Fig.\ \ref{fig.histAdjMOG_RAR}, yields the histograms of Fig.\
\ref{fig.histS133F}.  The best fit value of $a_0$ is nearly unchanged
from Fig.\ \ref{fig.histAdjMOG_RAR}.  The residuals of the MOG
histogram in Fig.\ \ref{fig.histS133F} are slightly increased, while
residuals of the RAR histogram are about the same as in the similar
histograms of Fig.\ \ref{fig.histAdjMOG_RAR}.  The great similarity of
the histograms of Figs.\ \ref{fig.histAdjMOG_RAR} and
\ref{fig.histS133F} indicates that the ability to find parameters that
give good MOG fits is not sensitive to the modeling assumptions.  The
equally good fit of the adjusted galaxies to the RAR curve, for either
set of assumptions, suggests that MOG predictions for galaxy rotation
are generically consistent with Eq.\ (\ref{Empiricalacc}).

\section{Supplementary Material}

Supplementary Material related to this article can be obtained
at~\url{https://doi.org/10.1016/j.dark.2019.100323}.

\vfill
\end{document}